\documentclass[twocolumn,aps,showpacs,amsmath,amssymb,superscriptaddress]{revtex4}

\usepackage{graphicx}
\usepackage{dcolumn}
\usepackage{bm}

\begin{document}


\title{Polariton-polariton scattering in microcavities: A microscopic theory}

\author{M. M. Glazov}
\affiliation{Ioffe Physical-Technical Institute, Russian Academy of Sciences, 194021 St. Petersburg, Russia}
\author{H. Ouerdane}
\affiliation{Centre de Recherche sur les Ions, les Mat\'eriaux et la Photonique, UMR CEA-CNRS-ENSICAEN-Universit\'e de Caen 6252, Boulevard Henri Becquerel, BP 5133, F-14070 Caen, France.}
\affiliation{LASMEA, UMR CNRS-Universit\'e Blaise Pascal 6602, 24 Avenue des Landais, 63177 Aubi\`ere Cedex France}
\author{L. Pilozzi}
\affiliation{Istituto dei Sistemi Complessi, CNR, I-00016, Cassella Postale 10, Monterotondo Stazione, Rome, Italy}
\author{G. Malpuech}
\affiliation{LASMEA, UMR CNRS-Universit\'e Blaise Pascal 6602, 24 Avenue des Landais, 63177 Aubi\`ere Cedex France}
\author{A. V. Kavokin}
\affiliation{School of Physics and Astronomy, University of Southampton, Highfield, Southampton SO17 1BJ, United Kingdom}
\affiliation{Marie-Curie Chair of Excellence ``Polariton devices'', University of Rome II, 1, via della Ricerca Scientifica, Rome, 00133, Italy}
\author{A. D'Andrea}
\affiliation{Istituto dei Sistemi Complessi, CNR, I-00016, Cassella Postale 10, Monterotondo Stazione, Rome, Italy}

\date{\today}

\begin{abstract}
We apply the fermion commutation technique for composite bosons to polariton-polariton scattering in semiconductor planar microcavities. Derivations are presented in a simple and physically transparent fashion. A procedure of orthogonolization of the initial and final two-exciton state wavefunctions is used to calculate the effective scattering matrix elements and the scattering rates. We show how the bosonic stimulation of the scattering appears in this full fermionic approach whose equivalence to the bosonization method is thus demonstrated in the regime of low exciton density. We find an additional contribution to polariton-polariton scattering due to the exciton oscillator strength saturation, which we analyze as well. We present a theory of the polariton-polariton scattering with opposite spin orientations and show that this scattering process takes place mainly via dark excitonic states. Analytical estimations of the effective scattering amplitudes are given.

\end{abstract}

\maketitle

\section{Introduction}

The study of exciton-polaritons (polaritons) is a rapidly developing area in modern condensed matter physics. Cavity polaritons, which are two-dimensional excitons strongly coupled to trapped photons in quantum microcavities~\cite{kavokin03b}, exhibit a rich variety of nonlinear effects. These include stimulated scattering~\cite{stim}, polarization rotation~\cite{rot}, bistable ~\cite{bistable} and multistable behaviors ~\cite{multistable}, superfluidity ~\cite{malpuech,yamamoto08}, and Bose condensation~\cite{bose,baumberg}. Because of these unique properties, cavity polaritons are among the promising candidates for the implementation of future low threshold optoelectronic devices and their investigation thus has become topical.

It is well established that the exciton-exciton interaction affects the exciton spin and momentum relaxation as well as the optical properties of semiconductors. The determination of the possible non-linearities governing optical properties is a fundamental question of prime importance in this field. Although it is quite clear that the excitonic interactions are the main source of nonlinearity, the specifics of microcavity polaritons has not been entirely revealed and the polarization dependence of exciton-exciton has not been described in detail. The main complication of the theoretical description of the exciton-exciton interaction arises from the fact that excitons are composite particles, each being formed of two fermions: an electron and a hole. At first glance, such a problem seems quite similar to that of interacting hydrogen atoms but the effective masses of positively and negatively charged carriers are comparable in semiconductors; therefore a direct generalization of the techniques developed in atomic physics for the study of hydrogen-hydrogen collisions~\cite{WOL93,JAM00,JAM09} to the problem of interacting excitons is not possible and this forbids any treatment of the exciton-exciton interaction by use of an effective Born-Oppenheimer (adiabatic) potential. Moreover, even though the exciton-exciton interaction energy is much smaller than the typical exciton binding energy, it cannot be treated by standard perturbation theory because of the electron-electron and hole-hole exchange processes. Indeed, it is impossible to know which electron is bound to which hole to form a given exciton and hence define the exciton-exciton interaction. This explains why much of the earlier attempts were based on oversimplified models neglecting exchange interaction between excitons~\cite{FEN87} or the exchange of carriers between excitons~\cite{KOH97}. This is not to mention the spin degree of freedom, which has been routinely neglected in the most part of previous works.

Studies of excitonic nonlinearities are usually performed by using essentially different strategies. One relies on the bosonization of excitons~\cite{HAN77,IVA98}, an approximation that was believed to hold at very low excitation density. In this approach, excitons are assumed to be the only constituents of the dilute system under consideration and they also are considered as bosonic elementary excitations experiencing an effective repulsion to avoid an overlap between the fermionic carriers wavefunctions. Under these assumptions, it was believed that by applying the Usui transformation~\cite{USU60} the fundamental Hamiltonian could be mapped to an effective bosonic Hamiltonian. Despite unresolved issues at the heart of the bosonic approach (because of very restrictive assumptions), its application has been justified by various groups~\cite{CIU98,TAS99,inoue00,BEN01,OKU01} on the grounds of the numerical results they obtain, which describe some features of experimental data \cite{AMA94,KUW97,AMA97,LEJ98,BUT99}, and under the assumption that inclusion of a two-body exciton interaction term in the bosonic Hamiltonian is sufficient to account for fermion exchange effects. More recently, an effective spin-dependent exciton-exciton interaction potential of the Heitler-London type including long-range van der Waals terms was generated and studied~\cite{SCH08}.

Another strategy is based on the full fermionic treatment of the problem, which amounts to solve the second-quantized equations of motion~\cite{LIN88}. Solutions of these equations rely on truncation schemes, which are tractable at the lowest order (Hartree-Fock), i.e. assuming the density of the electron-hole system to be high enough to neglect excitonic correlations owing to the screening of the Coulomb interaction. However, at lower densities, many-particle correlations have to be taken into account and the infinite hierarchy of equations satisfied by multipoint correlation functions should be truncated at higher orders, which in practice proves quickly intractable. This technical difficulty originates in the choice of the strength of the Coulomb interaction between carriers as the ``natural'' physical quantity to characterize and study the many-body problem in the interacting electron-hole system. In fact the Coulomb interaction cannot be treated as a perturbation when dealing with bound exciton states. Moreover, the random-phase approximation, extensively applied to this type of problems, amounts to factorize the multipoint correlation functions into two-point functions, neglecting in an uncontrollable way higher order Coulomb correlations such as biexcitons.

The optical excitation can also be considered as a pertinent physical parameter for a microscopic theory accounting for all many-body correlations in the photoexcited electron-hole system. Progress in the treatment of the equations of motion thus was made when a \emph{controlled} truncation scheme based on powers of the optical excitation strength was developped \cite{AXT94a,AXT94b,VIC95}. For instance, in the contexts of four-wave mixing and pump-probe experiments in the low excitation regime, the optical nonlinearities and related correlation effects are sufficiently well described at the third order of the applied laser field ($\chi^{(3)}$ response). This method, called the dynamics-controlled truncation (DCT) scheme, allows to derive a set of closed equations of motion describing the dynamics of the exciton-exciton interaction at the mean field level \emph{and} the four-body effects beyond mean field in a consistent way considering classical external laser fields \cite{OST95,OST98}. It also was succesfully applied to the study of six-wave mixing experiments with evidence of contributions to the signals of $\chi^{(5)}$ and $\chi^{(7)}$ processes \cite{AXT01}. The method was further improved with the inclusion of quantized electromagnetic fields in the Hamiltonian and the description of polaritonic effects \cite{SAV96,SCH07,POR08}.

The DCT scheme also provided a convenient framework for the analysis of biexcitonic correlations in terms of $T$-matrix \cite{TAK02}. More precisely, authors of Ref. \cite{TAK02} investigated the $\chi^{(3)}$ optical response of a semiconductor quantum well to relate the many-body effects to the exciton scattering amplitude. However, in many cases such as, e.g., non-resonant pumping of the system, it is more convenient to work on the basis of excitons or polaritons by introducing their effective scattering amplitudes. This approach proved extremely productive for the description of polarization-dependent kinetic phenomena in microcavities~\cite{kavokin03b,rot,inv}.

The problem of the scattering of two composite excitons and related exciton-exciton interaction has also been addressed quite recently in the series of works by M. Combescot and co-workers~\cite{COM02a,COM02b,COM04,COM07,COM08}, who developed a new technique based on commutation rules for the composite-exciton operators. Emphasizing the impossibility to define an interaction potential between two excitons because of the indistinguishability of their constituent carriers, their formalism was applied to study the scattering rates of excitons and certain differences with previous results were found~\cite{COM04}. The same group has considered the polariton-polariton scattering in microcavities and proposed a new type of the optical nonlinearity~\cite{COM07b}.

The present work is focused on the detailed theoretical description of polariton-polarion interactions in quantum microcavities. In Sec. II, we revisit the technique of Ref.~\cite{COM08} and present a somewhat simpler and physically more transparent description of exciton-exciton scattering. We demonstrate how the bosonic stimulation of exciton-exciton (or polariton-polariton) scattering can be obtained in Sec. III. We calculate effective scattering matrix elements for exciton-polaritons in microcavities and show how the previous results~\cite{TAS99} can be recovered. Special attention is paid to the nonlinearity caused by the saturation of the exciton oscillator strength in Sec. IV.

Since an exciton-polariton possesses a spin degree of freedom, it is characterized by the projection of the total angular momentum of the electron-hole pair, $+1$ or $-1$, on the growth axis. In Sec. V, we investigate in detail the effect of spin on the interaction of polaritons: it turns out that the scattering efficiency is strongly dependent on the mutual spin orientation of exciton-polaritons. We also present a microscopic derivation of the effective matrix elements for the scattering of polaritons with opposite spins, which is crucial for the linear polarization inversion observed in microcavities~\cite{inv}. In Sec. VI, we present analytical estimations of the effective scattering rates.

\section{Exciton-exciton scattering}\label{sec:basic}

In this section, we derive the basic quantities needed to calculate the scattering rates of composite bosons. To be specific, we consider excitons in a direct band-gap semiconductor quantum well. The splitting of the light-hole and heavy-hole bands is assumed to be large enough to neglect the population of the light-hole states. We first consider quantum well ground state excitons composed of electrons and heavy holes, ignoring their spin degree of freedom. The scattering rates derived here thus correspond to the scattering in the parallel-spin configuration. The scattering of excitons and exciton-polaritons with allowance for their spin is discussed in Sec.~\ref{sec:spin}.

\subsection{Wavefunctions}\label{sub:wavefunction}

Let $\varphi_i(\bm r_e, \bm r_h)$ be the wavefunction of a single exciton in the state $i$, where $\bm r_e$, $\bm r_h$ are the electron and hole two-dimensional position vectors. The state index $i$ is a global index, which accounts for both the states of the relative motion of an electron and a hole in the exciton (such as $1s, 2s,\ldots$) and the quantum numbers of the center of mass motion, with center of mass wavevector $\bm K$. We assume that the set of functions $\varphi_i$ forms an orthonormal basis:

\[
\int \varphi_j^*(\bm r_e, \bm r_h)\varphi_i(\bm r_e, \bm r_h) \mathrm d\bm r_e \mathrm d\bm r_h=\delta_{ij}.
\]

In the second quantization framework, the exciton creation operator $B_i^\dag$ can be defined as

\begin{equation}\label{Bdag}
B_i^\dag = \sum_{\bm k_e,\bm k_h} \tilde\varphi_i(\bm k_e, \bm k_h) a^\dag_{\bm k_e} b^\dag_{\bm k_h},
\end{equation}

\noindent where $\tilde\varphi_i(\bm k_e, \bm k_h)$ is the Fourier transform of $\varphi_i$, and $a^\dag_{\bm k_e}$ and $b^\dag_{\bm k_h}$ are the creation operators of the electron and hole with wavevectors $\bm k_e$ and $\bm k_h$ respectively. The sequential action of the operators $B_i^\dag$ on the vacuum state $|{\rm vac}\rangle$ creates the corresponding numbers of excitons in the state $i$. For instance, the two-exciton wavefunction

\begin{equation}\label{2ex}
\Psi_{ij}(\bm r_{e_1}, \bm r_{e_2},\bm r_{h_1}, \bm r_{h_2}) = B^\dag_{i} B^\dag_{j} |{\rm vac}\rangle,
\end{equation}

\noindent describes the pair of excitons in the states $i$ and $j$. The wavefunctions obtained in this manner are properly antisymmetrized products of single exciton envelopes and can be considered as a zero-order approximation to the exact two-exciton states described by the Hamiltonian:

\begin{equation}
\label{hamiltonian}
H_{\rm exc} = \sum_{a_l}T_{a_l}+{\sum_{a_l,b_{l'}}}'V_{a_l b_{l'}}(\bm r_{a_l} - \bm r_{b_{l'}}),
\end{equation}

\noindent where $T_{a_l}$ $(a = e,h; l = 1,2)$ are the kinetic energy operators of electrons and holes, and $V_{a_l b_{l'}}$ $(a,b = e,h; l,l' = 1,2)$ the electron-electron, electron-hole and hole-hole Coulomb interactions energies. The notation ${\sum}'$ in Eq. \eqref{hamiltonian} means that the terms with $l=l'$ corresponding to the same particles ($a=b$) should be excluded.

In systems of interacting excitons, the basis of two (and more) excitonic states is overcomplete and hence non-orthogonal. Assume that there are $N$ electron states and $N$ hole states in the system. The total number of excitonic states is $N_{exc} = N^2$ (as one can take an electron and a hole in any of possible states). Now, let us calculate the total number of 4-particle states considering two electrons and two holes. For a pair of electrons (holes) one easily finds $M = N(N-1)/2$ states as two fermions cannot occupy the same state. Thus, the number of 4-particle states is $M^2 \sim N^4/4$ for large $N$. However, the number of states with a pair of excitons defined by Eq. \eqref{2ex} is $N_{exc}^2/2 \sim N^4/2$ since pairs $(i,j)$ and $(j,i)$ are equivalent. In a general case, all these pair states $(i,j)$ have non-zero wavefunctions. Therefore, the number of two-exciton states is approximately twice as larger as the number of 4-particle states. Thus, a two-fold action of the operator $B^\dag$ on the vacuum state generates a set of mutually dependent wavefunctions, and this yields two important consequences: (i) in order to calculate any transition rate one should \emph{orthogonolize} the wavefunctions; (ii) the total lifetime of an exciton due to the exciton-exciton interaction will \emph{not} be equal to the sum of the transition rates of all possible two-exciton states.

\subsection{Effective matrix element of exciton-exciton scattering}\label{sub:matrix}

It is well established that excitonic nonlinearities in semiconductors are mainly caused by the exchange scattering of electron-hole complexes while the direct Coulomb interaction plays minor role. The exchange interaction has a short-range character (i.e. it becomes significant as the distance between the excitons' centers of mass becomes comparable to the excitonic Bohr radius, $a_B$). It means that e.g. for a two-dimensional system the exchange-induced corrections for a pair of excitons are governed by a small dimensionless parameter $\nu = a_B^2/S\ll 1$, where $S$ is the surface area of the sample. If the exciton gas is described by its surface density $n$, the parameter $\nu = na_B^2$. 

The overcompleteness of pair of excitons basis formed with the wavefunctions $\Psi(\bm r_{e_1}, \bm r_{e_2},\bm r_{h_1}, \bm r_{h_2})$,  defined in Eq.~\eqref{2ex}, as basis vectors, calls for a special procedure to calculate exciton-exciton scattering rates. Let $|i\rangle$, $|f\rangle$ be the initial and final two-exciton states, described by Eq.~\eqref{2ex}. In general, these states are not orthogonal and we denote their scalar product as

\begin{equation}\label{c}
\mathcal C = \langle i|f\rangle.
\end{equation}

\noindent Physically, it is obvious that the transitions should be considered only between the orthogonal states; otherwise part of the final state is admixed to the initial state. As we shall see below the constant $\mathcal C$ is proportional to the parameter $\nu$ and hence small. Therefore one may seek to consider orthogonolized pairs of states as slightly modified initial and final states, namely,

\begin{equation}
|i'\rangle = |i\rangle - \alpha |f\rangle, \quad |f'\rangle = |f\rangle - \beta |i\rangle.
\end{equation}

\noindent where $\alpha$ and $\beta$ are some small complex coefficients ($\alpha, \beta \propto \nu$). The scalar product of the states $|i'\rangle$, $|j'\rangle$ reads

\begin{equation}
\langle i'|f'\rangle = \mathcal C - \alpha^*\langle f|f \rangle - \beta\langle i|i \rangle.
\end{equation}

\noindent The quadratic terms in $\alpha$ and $\beta$ are neglected and the newly defined states are thus orthogonal provided that $\alpha^*\langle f|f \rangle+\beta\langle i|i \rangle = \mathcal C$ (the asterisk symbol denotes the complex conjugation). 

The matrix element of the total Hamiltonian between the orthogonalized states is

\begin{equation}\label{ort1}
\langle f'| H_{\rm exc}| i' \rangle = \langle f| H_{\rm exc}| i \rangle - \alpha \langle f| H_{\rm exc}|f\rangle - \beta^*\langle i| H_{\rm exc}|i\rangle.
\end{equation}

Note that in the resonant situation where the energies of the initial and final states are equal, the matrix elements $\langle f| H_{\rm exc}|f\rangle/\langle f|f \rangle$ and $\langle i| H_{\rm exc}|i\rangle/\langle i|i \rangle$ are equal. From Eq. \eqref{ort1} we obtain the transition matrix element $\mathcal M_{i\to f}$ as

\begin{equation}\label{Mif}
\mathcal M_{i\to f} = \langle f'| H_{\rm exc}| i' \rangle = \langle f| H_{\rm exc}| i \rangle - \mathcal C^* \frac{\langle i| H_{\rm exc}|i\rangle}{\langle i|i \rangle},
\end{equation}

\noindent which is independent of the orthogonalization procedure, provided that the parameter $\nu$ is small enough. Note, that in a general case the second term in Eq.~\eqref{Mif} is not small as compared with the first one. Moreover, if one can formally separate the interaction part of the Hamiltonian, Eq.~\eqref{Mif} immediately gives the correct value for the transition matrix element as the matrix element of the interaction part.

In order to establish the link between Eq.~\eqref{Mif} and the results of Ref.~\cite{COM08} we calculate the transition matrix element in a simple case: two excitons initially occupy the same quantum state $|i\rangle = |00\rangle$ while the final states for a pair of particles are defined by $|f\rangle = |12\rangle = |21\rangle$. It is assumed that the conservation of energy between this states is fulfilled. To proceed with the calculations the knowledge of three quantities is needed: the scalar product of the initial and final states, the matrix element of the full Hamiltonian between the initial and final states, and the mean value of the energy in the initial (or final) state. 

The scalar product of the states $|i\rangle$ and $|f\rangle$ given by:

\begin{eqnarray}\label{scalar_prod}
\mathcal C = \langle f | i\rangle = \langle 12 | 00 \rangle & = & \langle{\rm vac}|B_1B_2B_0^\dag B_0^\dag|{\rm vac}\rangle
\nonumber\\
& = & -2\lambda\left(
\begin{array}{cc}
2 & 0 \\
1 & 0 \\
\end{array}\right),
\end{eqnarray}

\noindent can be derived using the commutation relations \cite{COM08} or by expansion of Slater determinants. The quantity $\lambda$ appearing in the equation above is the dimensionless Pauli parameter \cite{COM07,COM08} that describes an overlap between the initial and final state. More precisely, it is given by:

\begin{align}\label{lambda}
\lambda\left(
\begin{array}{cc}
n & j \\
m & i \\
\end{array}\right) = 
\int \mathrm d\bm r_{e_1}\mathrm d\bm r_{e_2} \mathrm d\bm r_{h_1}\mathrm d\bm r_{h_2} \\ \varphi_m^*(\bm r_{e_1}, \bm r_{h_2})\varphi_n^*(\bm r_{e_2}, \bm r_{h_1})\varphi_i(\bm r_{e_1}, \bm r_{h_1})\varphi_j(\bm r_{e_2}, \bm r_{h_2}) . \nonumber
\end{align}

\noindent Equation~\eqref{scalar_prod} agrees with Eq. (7.4) of Ref.~\cite{COM08}. It is worth noting that the integral in Eq.~\eqref{scalar_prod} describes the probability for two excitons to be at the same place and its magnitude is thus on the order of $\nu$.

The matrix element of the Hamiltonian between the states $|i\rangle$ and $|f\rangle$ reads

\begin{eqnarray}\label{h_elem}
\langle f| H_{\rm exc}| i \rangle & = & \langle{\rm vac}| B_1B_2  H_{\rm exc} B_0^\dag B_0^\dag|{\rm vac}\rangle
\nonumber\\
& = &-4E_0\lambda\left(
\begin{array}{cc}
1 & 0 \\
2 & 0 \\
\end{array}\right)
\nonumber\\
&& + 2\xi\left(
\begin{array}{cc}
1 & 0 \\
2 & 0 \\
\end{array}
\right) 
- 2\xi^{in}\left(
\begin{array}{cc}
1 & 0 \\
2 & 0 \\
\end{array}\right),
\end{eqnarray}

\noindent where $E_0$ is the energy of an exciton in the state $0$. Here the quantities $\xi$ and $\xi^{in}$, which respectively describe the Coulomb direct scattering, and the Coulomb exchange scattering between the \emph{in} excitons, i.e. excitons in the states $i$ and $j$ in the terminology of Refs.~\cite{COM07,COM08}, are defined as follows:

\begin{widetext}
\begin{eqnarray}\label{xi}
\xi\left(
\begin{array}{cc}
n & j \\
m & i \\
\end{array}
\right) 
&=& \int \mathrm d\bm r_{e_1}\mathrm d\bm r_{e_2} \mathrm d\bm r_{h_1}\mathrm d\bm r_{h_2} \varphi_m^*(\bm r_{e_1}, \bm r_{h_1})\varphi_n^*(\bm r_{e_2}, \bm r_{h_2})\varphi_i(\bm r_{e_1}, \bm r_{h_1})\varphi_j(\bm r_{e_2}, \bm r_{h_2})
\nonumber\\
&&\times\left[V_{ee}(\bm r_{e_1} - \bm r_{e_2}) +V_{hh}(\bm r_{h_1} - \bm r_{h_2}) + V_{eh}(\bm r_{e_1} - \bm r_{h_2}) + V_{eh}(\bm r_{e_2} - \bm r_{h_1}) \right],
\end{eqnarray}

\noindent and

\begin{eqnarray}\label{xi_in}
\xi^{in}\left(
\begin{array}{cc}
n & j \\
m & i \\
\end{array}\right) 
& = & \int \mathrm d\bm r_{e_1}\mathrm d\bm r_{e_2} \mathrm d\bm r_{h_1}\mathrm d\bm r_{h_2} \varphi_m^*(\bm r_{e_1}, \bm r_{h_2})\varphi_n^*(\bm r_{e_2}, \bm r_{h_1})\varphi_i(\bm r_{e_1}, \bm r_{h_1})\varphi_j(\bm r_{e_2}, \bm r_{h_2})
\nonumber\\
&&\times\left[V_{ee}(\bm r_{e_1} - \bm r_{e_2}) +V_{hh}(\bm r_{h_1} - \bm r_{h_2}) + V_{eh}(\bm r_{e_1} - \bm r_{h_2}) + V_{eh}(\bm r_{e_2} - \bm r_{h_1}) \right].
\end{eqnarray}
\end{widetext} 

\noindent Finally, we need to calculate the mean value of the Hamiltonian in the state $|i\rangle$. We find:

\begin{equation}\label{H_aver}
\langle i| H_{\rm exc}|i\rangle \approx 4E_0.
\end{equation}

Combining Eqs. \eqref{scalar_prod}, \eqref{h_elem} and \eqref{H_aver} we obtain

\begin{equation}\label{Mif1}
\mathcal M_{i \to f} = 2\xi\left(
\begin{array}{cc}
2 & 0 \\
1 & 0 \\
\end{array}
\right)
- 2\xi^{in}\left(
\begin{array}{cc}
2 & 0 \\
1 & 0 \\
\end{array}
\right).
\end{equation}

\subsection{Transition rate}\label{sub:transition}

Now we calculate the transition rate between the initial and final states. To this end, we use the Fermi golden rule and take into account that the states $|i\rangle$ and $|f\rangle$ are not normalized. The transition rate reads

\begin{equation}\label{transition_rate}
\frac{1}{\tau_{i\to f}} = \frac{2\pi}{\hbar} \frac{|\mathcal M_{i \to f}|^2}{\langle i|i\rangle\langle f|f\rangle}~ \delta(2E_0 - E_1 - E_2).
\end{equation}

Since the matrix element ${\mathcal M}_{i \to f}$ is proportional to the small quantity $\nu$, one can neglect the composite nature of excitons in the calculation of the normalization constants:
$\langle f|f\rangle = \langle 12|12\rangle \approx 1$
and
$\langle i|i\rangle = \langle 00|00\rangle \approx 2$ [also used to derive Eq.~\eqref{Mif1}].
For the same reason one can use free exciton energies in the energy-conservation law. As a result we obtain:

\begin{equation}\label{transition_rate1}
\frac{1}{\tau_{i\to f}} = \frac{4\pi}{\hbar} \left|\xi\left(
\begin{array}{cc}
2 & 0 \\
1 & 0 \\
\end{array}\right) 
- \xi^{in}\left(
\begin{array}{cc}
2 & 0 \\
1 & 0 \\
\end{array}
\right)\right|^2 
\delta(2E_0 - E_1 - E_2),
\end{equation}

\noindent which is equivalent to Eq. (11.18) of \cite{COM08} and Eq. (13) of Ref.~\cite{COM04}, and we have thus proven that our simplified approach based on the orthogonalization of the initial and final states yields the same result for the exciton-exciton scattering rate as the more elaborate approach of Combescot and co-workers \cite{COM08}.

\section{Bosonic stimulation induced by exciton-exciton scattering}\label{sec:bosonic}

In the framework of the fermionic approach revisited here an important question arises as whether it is possible to reproduce correctly the bosonic effects, mainly the bosonic stimulation of the exciton-exciton scattering.

Here we show how to deduce the bosonic stimulation from the study of the exciton-exciton scattering matrix element. Following the approach developed in Ref.~\cite{COM08}, we consider the simplest case of a single exciton in the final state, thus the expected enhancement factor of the inverse transition rate $\tau_{i\to f}^{-1}$ is equal to 2. The problem is complex because one needs to consider not two, but \emph{three}-particle states.

The initial and final states are given by:

\begin{equation}\label{bos:if}
|i\rangle = |100\rangle = B_1^\dag(B_0^\dag)^2|{\rm vac}\rangle, \quad |f\rangle = |211\rangle = B_2^\dag(B_1^\dag)^2|{\rm vac}\rangle.
\end{equation}

\noindent It is worth noting that, as in Sec.~\ref{sub:matrix}, these states are neither orthogonal, nor normalized. As for the normalization constants we need only the main contribution:

\begin{equation}\label{bos:norm}
\langle i|i\rangle \approx \langle f|f\rangle \approx 2.
\end{equation}

The calculation of the scalar products of the states $|i\rangle$ and $|f\rangle$ is more complex. As shown in Appendix~\ref{app:scal}, we obtain the scalar product $\langle f|i\rangle$ to the lowest (non-vanishing) order in $\nu$ as:

\begin{equation}\label{bos:scalar}
\langle f|i\rangle = \langle{\rm vac}|B_1^2B_2 (B_0^\dag)^2B_1^\dag|{\rm vac}\rangle \approx -4\lambda\left(
\begin{array}{cc}
1 & 0 \\
2& 0 \\
\end{array}
\right).
\end{equation}

The matrix element of the total Hamiltonian between the initial and final states reads

\begin{equation}\label{bos:h0}
\langle f| H_{\rm exc}|i\rangle = \langle{\rm vac}|B_1^2B_2  H_{\rm exc} (B_0^\dag)^2B_1^\dag|{\rm vac}\rangle.
\end{equation}

\noindent Rather tedious algebra (see Appendix~\ref{app:ham}) yields the leading order contributions to Eq.~\eqref{bos:h0}:

\begin{equation}\label{bos:h0:1}
\langle f| H_{\rm exc} |i \rangle \approx d + (E_1 + 2E_0) \langle f| i \rangle,
\end{equation}
where
\begin{equation}\label{bos:d21}
d = 4
\xi\left(
\begin{array}{cc}
2 & 0 \\
1 & 0 \\
\end{array}\right) 
- 4\xi^{in}\left(
\begin{array}{cc}
2 & 0 \\
1 & 0 \\
\end{array}
\right)
\end{equation} 

\noindent is proportional to $\nu$.

In agreement with Eq.~\eqref{Mif} the transition matrix element reads

\begin{equation}\label{bos:m}
\mathcal M_{i\to f} = \langle f| H_{\rm exc} |i \rangle - \langle f| i \rangle \frac{\langle i| H_{\rm exc}| i \rangle}{\langle i|i \rangle} \approx d,
\end{equation}

\noindent and the exciton transition rate can be recast as

\begin{equation}\label{bos:transition_rate1}
\frac{1}{\tau_{i\to f}} = 2\frac{4\pi}{\hbar} \left|\xi\left(
\begin{array}{cc}
2 & 0 \\
1 & 0 \\
\end{array}\right) 
- \xi^{in}\left(
\begin{array}{cc}
2 & 0 \\
1 & 0 \\
\end{array}
\right)\right|^2 \delta(2E_0 - E_1 - E_2),
\end{equation}

\noindent which is exactly twice larger than the transition rate with unoccupied final state, Eq.~\eqref{transition_rate1}. 

Equation~\eqref{bos:transition_rate1} clearly demonstrates that the fermionic approach yields, to the leading order in $\nu$, the bosonic enhancement of the scattering rate. It is worth noting that the corrections at the next order in $\nu$ reduce the transition rate as compared with Eq.~\eqref{bos:transition_rate1}. In the case of $N$ excitons in the final state, the procedure described above leads to the $N+1$ fold increase of the transition rate provided $N\nu \ll 1$.

The transitions between fixed set of initial and final states thus can be described in the `weakly-interacting' bosons approximation. We note, however, that such a procedure cannot be applied to the case where the transitions cover all range of possible final states. Indeed, in order to calculate the exciton lifetime with respect to the transitions to all final states one needs to orthogonalize not only the initial and final states, but also all the final states between themselves. Since the basis of the pair of exciton states is overcomplete, it yields an extra factor of $2$ in the transition rate, as shown in Ref.~\cite{COM04} 

\section{Polariton-polariton scattering}

Methods developed in Sec. II and III are applied here to the case of exciton-polaritons in semiconductor planar microcavities. Below we briefly review the basics of light-matter coupling in planar microcavities and afterwards calculate polariton-polariton effective scattering matrix elements due to (i) exciton-exciton scattering and (ii) the nonlinearity resulting from the saturation of the light-matter interaction strength.

\subsection{Polariton states in quantum microcavities}

Consider the interaction between electron-hole pairs with the light field. In the dipole approximation the formation of an exciton after the absorption of a photon is described by the dipole transition operator which can be conveniently expressed in terms of electron and hole creation operators as:

\begin{equation}\label{polarization_k1}
P^\dag_{\bm \varkappa} = \sum_{\bm k_e, \bm k_h} a_{\bm k_e}^\dag b_{\bm k_h}^\dag \delta_{\bm k_e + \bm k_h,\bm \varkappa},
\end{equation}

\noindent where $\bm \varkappa$ denotes the photon wavevector. The requirement of momentum conservation in the process of photon absorption is satisfied by the presence of the Kronecker $\delta$. Note that although the absolute value of the wavevector $\bm \varkappa$ involved in an optical transition is much smaller than those of the electron and hole wavevectors $k_e, k_h (\sim a_B^{-1})$, $\bm \varkappa$ cannot be neglected here in order to ensure the coupling of the photon and exciton with the same in-plane wavevectors. The light-matter interaction can thus be described by the following Hamiltonian

\begin{equation}\label{Hlmc}
H_{\rm exc-ph} = \sum_{\bm \varkappa} (W P^\dag_{\bm \varkappa} A_{\bm \varkappa} + W^* P_{\bm \varkappa} A_{\bm \varkappa}^\dag),
\end{equation}

\noindent where $A_{\bm \varkappa}^\dag$ is the photon creation operator and $W$ is a constant defined below. Note that the operators $A_{\bm \varkappa}^\dag$ and $A_{\bm \varkappa}$ are \emph{true} bosonic operators.

In Ref.~\cite{COM07b}, $H_{\rm exc-ph}$ is defined with a light-matter coupling \emph{linear} in the exciton operators:

\begin{equation}\label{Hlmc2}
H_{\rm exc-ph}^{\rm lin} = \sum_{\bm \varkappa} (\Omega_R B^\dag_{\bm \varkappa} A_{\bm \varkappa} + \Omega_R^* B_{\bm \varkappa} A_{\bm \varkappa}^\dag),
\end{equation}

\noindent where $\Omega_R$ denotes the Rabi splitting, characterizing the linear coupling between a photonic mode and an excitonic state. Hereafter we consider only $1s$ excitons and enumerate these states in accordance with their in-plane momentum. One can check that both Hamitonians, in Eqs.~\eqref{Hlmc} and \eqref{Hlmc2}, have the same single-particle eigenstates if the two constants $W$ and $\Omega_R$ satisfy

\[
 \Omega_R = W \int \varphi_{\bm \varkappa}^*(\bm r, \bm r) e^{{-\mathrm i \bm \varkappa \cdot \bm r}}\mathrm d\bm r.
\]

The cavity polaritons are formed as coherent superpositions of the exciton states in a quantum well and photon states in a planar microcavity. In what follows we consider only polaritons arising due to strong coupling of the single cavity mode characterized by an in-plane wavevector $\bm \varkappa$ with the $1s$ heavy-hole exciton ground state with an in-plane wavevector $\bm K$. The translational invariance in the plane of the structure imply the equality of the wavevectors of exciton and photon constituting a polariton, $\bm K = \bm \varkappa$. Moreover, we consider the states on the lower polariton branch only, therefore exciton-polaritons can be enumerated by a wavevector $\bm \varkappa$ and their creation operator can be written as

\begin{equation}\label{polariton_i}
C^\dag_{\bm \varkappa} = \alpha_{\bm \varkappa} A_{\bm \varkappa}^\dag + \beta_{\bm \varkappa} B_{\bm \varkappa}^\dag,
\end{equation}

\noindent where unimportant extra quantum numbers are omitted for ease of notations. The prefactors $\alpha_{\bm \varkappa}$ and $\beta_{\bm \varkappa}$ are Hopfield coefficients.

The lower branch polariton dispersion reads:

\begin{eqnarray}\label{LPdisp}
E_{\bm \varkappa}^{P} & = & \frac{1}{2} \left(\delta + \frac{\hbar^2\varkappa^2}{2M_X} + \frac{\hbar^2\varkappa^2}{2m_{ph}}\right)
\nonumber\\
&&-\frac{1}{2} \sqrt{\left(\delta + \frac{\hbar^2\varkappa^2}{2M_X} - \frac{\hbar^2\varkappa^2}{2m_{ph}}\right)^2 + \Omega_R^2}~,
\end{eqnarray}

\noindent where $\delta$ is the detuning between the photon and the exciton modes, $M_X$ is the exciton total mass, and $m_{ph}$ is the photon effective mass, which depends on the width of the cavity and the refractive index. Here we have neglected the renormalization of the expression under the square root due to the broadening of the exciton and photon modes. Note that the Hopfield coefficients can be directly linked to the photon and exciton energies through:

\begin{equation}
\frac{2\alpha_{\bm \varkappa}\beta_{\bm \varkappa}}{\alpha_{\bm \varkappa}^2-\beta_{\bm \varkappa}^2} = \Omega_R \left(\frac{\hbar^2\varkappa^2}{2m_{ph}} - \delta - \frac{\hbar^2\varkappa^2}{2M_X} \right)^{-1}.
\end{equation}

\subsection{Effective matrix elements and polariton-polariton scattering rates}

Here we consider two main contributions to the polariton-polariton scattering cross section. One of them results from the exciton-exciton scattering, see Sec. \ref{sec:basic}; the other one has its origins in the nonlinearity of the light-matter Hamiltonian, Eq.~\eqref{Hlmc}. Below we calculate these two contributions, and we start by defining the two-polariton states as $|\bm \varkappa, \bm \varkappa'\rangle = C_{\bm \varkappa}^\dag C_{\bm \varkappa'}^\dag |{\rm vac}\rangle$, which we write explicitely as:

\begin{multline}\label{2pol}
|\bm \varkappa, \bm \varkappa'\rangle = \left(\alpha_{\bm \varkappa}\alpha_{\bm \varkappa'} A_{\bm \varkappa}^\dag A_{\bm \varkappa'}^\dag + \beta_{\bm \varkappa}\beta_{\bm \varkappa'} B_{\bm \varkappa}^\dag B_{\bm \varkappa'}^\dag\right. \\ + \left. \alpha_{\bm \varkappa} \beta_{\bm \varkappa'} A_{\bm \varkappa}^\dag B_{\bm \varkappa'}^\dag + \alpha_{\bm \varkappa'} \beta_{\bm \varkappa} A_{\bm \varkappa'}^\dag B^\dag_{\bm \varkappa}\right) |{\rm vac}\rangle.
\end{multline}

For simplicity of analysis, we consider the scattering of two polaritons initially in the same state $|i\rangle \equiv |\bm \varkappa_0\bm \varkappa_0\rangle$ to the final states $|f\rangle \equiv |\bm \varkappa_1\bm \varkappa_2\rangle$. Such a situation corresponds to the optical parametric oscillators based on microcavities, which are widely studied both experimentally and theoretically, see e.g. Refs.~\cite{stim,SAB01,CIU01,CIU03}. In what follows we omit $\bm \varkappa$ in all places where it does not lead to confusion. The overlap of these states is 

\begin{equation}\label{dot_prod}
\langle f|i\rangle = \beta_1^*\beta_2^*\beta_0^2 \langle{\rm vac}|B_1 B_2 (B_0^\dag)^2|{\rm vac}\rangle = 
-2\beta_1^*\beta_2^*\beta_0^2 \lambda 
\left(
\begin{array}{cc}
1 & 0 \\
2 & 0 
\end{array}
\right),
\end{equation}

\noindent It is worth noting that the non-orthogonality of the polariton states is determined by the non-orthogonality of the exciton states because photons are true bosons.

The total Hamiltonian of the system $H_{\rm tot}$ is the sum of three contributions: the excitonic Hamiltonian, $ H_{\rm exc}$, given in Eq.~\eqref{hamiltonian}, the light-matter coupling Hamiltonian, $H_{\rm exc-ph}$, given in Eq.~\eqref{Hlmc}, and the photonic Hamiltonian,

\begin{align}
\label{h_tot_pol}
H_{\rm tot} = \sum_{a_l}T_{a_l}+{\sum_{a_l,b_{l'}}}'V_{a_l b_{l'}}(\bm r_{a_l} - \bm r_{b_{l'}}) + \nonumber\\
\sum_{\bm \varkappa} (W P^\dag_{\bm \varkappa} A_{\bm \varkappa} + W^* P_{\bm \varkappa} A_{\bm \varkappa}^\dag)+ \sum_{\bm \varkappa} E^{ph}_{\bm \varkappa} A^\dag_{\bm \varkappa} A_{\varkappa}
\end{align}

The matrix element of the total Hamiltonian $H$ between the states $|i\rangle$ and $|j\rangle$ can be recast as a sum of two terms: one is related to the exciton-exciton interaction and is given, in accord with Eq.~\eqref{h_elem}, by

\begin{eqnarray}
\nonumber
\beta_1^*\beta_2^*\beta_0^2
\left[ 2\xi
\left(
\begin{array}{cc}
2 & 0 \\
1 & 0 \\
\end{array}
\right)
 - 2\xi^{in}
\left(
\begin{array}{cc}
2 & 0 \\
1 & 0 \\
\end{array}
\right)
-4E_0^X \lambda \left( 
\begin{array}{cc}
 2 & 0\\
1 & 0
\end{array}
\right)\right] .
\end{eqnarray}

The other contribution arises from the light-matter interaction part of the Hamiltonian and is related to the nonlinearity in the optical absorbtion caused by the saturation of exciton oscillator strength. It is also called ``\emph{photon-mediated exchange}'' ~\cite{COM07b}. The calculation of the matrix element $\langle f | H_{\rm exc-ph}| i \rangle$ yields the following result:

\begin{subequations}
\begin{eqnarray}
\langle f | H_{\rm exc-ph}| i \rangle = \nonumber \\
\alpha_1^*\beta_2^* \beta_0^2 \langle{\rm vac}| A_1 B_2  H_{\rm exc-ph} (B_0^\dag)^2 |{\rm vac}\rangle + \label{1} \\
\alpha_2^*\beta_1^* \beta_0^2 \langle{\rm vac}| A_2 B_1  H_{\rm exc-ph} (B_0^\dag)^2 |{\rm vac}\rangle + \label{2}\\
2\beta_1^*\beta_2^* \alpha_0\beta_0 \langle{\rm vac}| B_1 B_2  H_{\rm exc-ph} A_0^\dag B_0^\dag |{\rm vac}\rangle \label{3}. 
\end{eqnarray}
\end{subequations}

The physical sense of these processes can be illustrated, for instance, in the case of term Eq.~\eqref{1}. The initial state of two polaritons $|00\rangle$ contains (with some weight) the product $B_0^\dag B_0^\dag|{\rm vac}\rangle$, but the pairs $B_0^\dag B_0^\dag|{\rm vac}\rangle$ and $B_1^\dag B_2^\dag|{\rm vac}\rangle$ are not orthogonal and hence one of these excitons can recombine and emit a photon $A_1^\dag|{\rm vac}\rangle$. After the emission, there is a probability for the remaining exciton to end up in the state 2. This state is nothing but a part of the polariton in $C_1^\dag C_2^\dag|{\rm vac}\rangle$. Such a process is allowed by the momentum and energy conservation law, provided that $2E_0^P = E_1^P + E_2^P$ and $2\bm \varkappa_0 = \bm \varkappa_1 + \bm \varkappa_2$. Using the identities given in appendix~\ref{app:d}, these matrix elements can be evaluated explicitly:

\begin{eqnarray}\label{pme}
&&\langle f | H_{\rm exc-ph}| i \rangle = -2 \alpha_1^*\beta_2^*\beta_0^2 \rho_{-\bm \varkappa_1}(0,0;2) W\\
\nonumber
&&- 2\alpha_2^*\beta_1^*\beta_0^2 \rho_{-\bm \varkappa_2}(0,0;1) W\\
\nonumber
&&- 2\beta_1^*\beta_2^*\beta_0\alpha_0 \left[\rho_{\bm \varkappa_0}(1,2;0)+ \rho_{\bm \varkappa_0}(2,1;0)\right] W^*.
\end{eqnarray}

\noindent where

\begin{equation}\label{rho}
\rho_{\bm \varkappa}(i,j;k) = \int \mathrm d \bm r \mathrm d \bm r_e \mathrm d \bm r_h \varphi^*_i(\bm r_e, \bm r) 
\varphi^*_j(\bm r, \bm r_h) \varphi_k(\bm r_e, \bm r_h) e^{\mathrm i \bm \varkappa \bm r}.
\end{equation}

In the same fashion as above for the exciton-exciton scattering, we take into account the non-orthogonality of the two-polariton states Eq.~\eqref{Mif} and use the modified Fermi golden rule Eq.~\eqref{transition_rate}, to obtain the polariton transition rate as:

\begin{widetext}
\begin{eqnarray}\label{transition_rate2}
&&\frac{1}{\tau_{i\to f}} = \frac{4\pi}{\hbar} \left|
\beta_1^*\beta_2^* \beta_0^2 \left\{ \xi\left(
\begin{array}{cc}
2 & 0 \\
1 & 0 \\
\end{array}\right) 
- \xi^{in}
\left(
\begin{array}{cc}
2 & 0 \\
1 & 0 \\
\end{array}\right)
\right\}
-\alpha_1^*\beta_2^*\beta_0^2 \rho_{-\bm \varkappa_1}(0,0;2) W - \alpha_2^*\beta_1^*\beta_0^2 \rho_{-\bm \varkappa_2}(0,0;1) W\right.
\\
\nonumber
&&\left.- 2\beta_1^*\beta_2^*\beta_0\alpha_0 \left[\rho_{-\bm \varkappa_0}(1,2;0)+ \rho_{-\bm \varkappa_0}(2,1;0)\right] W^*
-\beta_1^*\beta_2^*\beta_0^2\lambda \left( 
\begin{array}{cc}
 2 & 0\\
1 & 0
\end{array}
\right) \left[\left(E_0^X - E_0^{ph}\right)|\alpha_0|^2  - \alpha_0\beta_0^*\Omega_R - \alpha_0^*\beta_0 \Omega_R^*\right]\right|^2
\\
&&\times \delta(2E_0^P - E_1^P - E_2^P).
\nonumber 
\end{eqnarray}
\end{widetext}

Equation~\eqref{transition_rate2} is the central result of this section. As compared with Ref.~\cite{COM07b} we use a different form of the light-matter coupling Hamiltonian, Eq.~\eqref{Hlmc}, which takes into account the fact that in the process of photon absorption the electron-hole pair is created in the same point in real space. The analogous procedure with an effective Hamiltonian, Eq.~\eqref{Hlmc2}, would yield the same result as published in Ref.~\cite{COM07b}. The main difference between our results and those of Ref.~\cite{COM07b} is in the appearance of the $\rho$ overlap integrals in our case, see Eq.~\eqref{rho}. These overlap integrals contain products of three excitonic wavefunctions and one photonic wavefunction. This implicitly accounts for the fact that, e.g., in the process of photon absorption along with the creation of an additional exciton, the ``resident'' exciton in microcavity may change its quantum state. The non-linearity due to the light-matter interaction has been discussed also in Refs.~\cite{TAS99,ROC00,CIU00}, where overlap integrals of the form Eq.~\eqref{rho} also appear. However, the effects of the non-orthogonality of the exciton wavefunctions on the effective matrix elements of polariton-polariton interaction were disregarded in these works.

To conclude this section, it is worth noting that the contributions to the effective matrix element of polariton-polariton scattering due to light-matter interaction both in Ref.~\cite{COM07b} and in Eqs.~\eqref{pme}, \eqref{transition_rate2} can be estimated as $\sim \nu \Omega_R$ (see Eq.~\eqref{final}). Thus, qualitatively, the difference in the approaches of Ref.~\cite{COM07b}, Refs.~\cite{TAS99,ROC00,CIU00} and ours merely yields different numerical prefactors before the scattering amplitudes, while their dependence on the system parameters remains essentially the same.

\section{Scattering of polaritons with opposite spin orientations}\label{sec:spin}

So far, the effect of spin on the polariton-polariton scattering was not discussed and our consideration was restricted to a fully spin-polarized case: all polaritons were assumed to have the same projection of the momentum onto the growth axis. Below we extend our treatment to allow for the spin degree of freedom of polaritons.

The spin projection of an electron on the microcavity growth axis is $\pm 1/2$ and the spin projection of a heavy hole is $\pm 3/2$. It is worth noting that the circularly polarized $\sigma^+$ excitation creates an electron-hole pair (or exciton) in the spin state $(e,hh) = (-1/2,3/2)$ which has a $z$-component of the total angular momentum $m_z=+1$. The $\sigma^-$ polarized excitation creates an exciton in the spin state $(1/2,-3/2)$ with $z$-component of the angular momentum being $m_z=-1$. The two remaining exciton states with $m_z = \pm 2$, namely, $(1/2,3/2)$ and $(-1/2,-3/2)$ are qualified as dark: they do not take part in the light-matter coupling but may serve as intermediate states for the scattering process.

\subsection{General remarks}

Here, we are interested in the situation where two excitons are created by two pump pulses with opposite circular polarizations, $\sigma^+$ and $\sigma^-$. In such a case, a pair of excitons whose total $z$-component of the angular momentum is equal to $0$, is created. Let us introduce the exciton creation operator in the state with $m_z=+1$, $B_{i,+1}^\dag$, as

\begin{equation}\label{Bdag_p}
B_{i,+1}^\dag = \sum_{\bm k_e,\bm k_h} \tilde\varphi_i(\bm k_e, \bm k_h) a^\dag_{\bm k_e,-1/2} b^\dag_{\bm k_h,3/2} ,
\end{equation}

\noindent where $\tilde\varphi_i(\bm k_e, \bm k_h)$ is the Fourier transform of the exciton envelope function $\varphi(\bm r_e, \bm r_h)$, $a^\dag_{\bm k_e,\pm 1/2}$, $b^\dag_{\bm k_h,\pm 3/2}$ are the electron and hole creation operators. Here it is assumed that the subscript $i$ denotes the orbital states of an exciton, say, $i=(1s,\bm K)$, and the second subscript denotes the $z$-component of the spin. Analogously, the exciton creation operator in the state with $m_z=-1$, $B_{i,-1}^\dag$, reads

\begin{equation}\label{Bdag_m}
B_{i,-1}^\dag = \sum_{\bm k_e,\bm k_h} \tilde\varphi_i(\bm k_e, \bm k_h) a^\dag_{\bm k_e,+1/2} b^\dag_{\bm k_h,-3/2}.
\end{equation}

\noindent We need also two more exciton states which are dark, namely, those with $m_z = \pm 2$. Their creation operators read

\begin{subequations}
\begin{equation}\label{dp}
B_{i,+2}^\dag = \sum_{\bm k_e,\bm k_h} \tilde\varphi_i(\bm k_e, \bm k_h) a^\dag_{\bm k_e,1/2} b^\dag_{\bm k_h,3/2},
\end{equation}
\begin{equation}\label{dm}
B_{i,-2}^\dag = \sum_{\bm k_e,\bm k_h} \tilde\varphi_i(\bm k_e, \bm k_h) a^\dag_{\bm k_e,-1/2} b^\dag_{\bm k_h,-3/2}.
\end{equation}
\end{subequations}

Let us now proceed with the pair of excitons states. Under the conditions described above the pairs of excitons are formed with opposite spin orientations:

\begin{subequations}
\begin{equation}\label{lpairs}
|i,+1;j,-1\rangle = B_{i,+1}^\dag B_{j,-1}^\dag |{\rm vac}\rangle,
\end{equation}
\begin{equation}
|i,-1;j,+1\rangle = B_{i,-1}^\dag B_{j,+1}^\dag |{\rm vac}\rangle,
\end{equation}
\begin{equation}
|i,+2;j,-2\rangle = B_{i,+2}^\dag B_{j,-2}^\dag |{\rm vac}\rangle,
\end{equation}
\begin{equation}
\ |i,-2;j,+2\rangle = B_{i,-2}^\dag B_{j,+2}^\dag |{\rm vac}\rangle.
\end{equation}
\end{subequations}
The following orthogonality relations greatly simplify our consideration:
\begin{subequations}
\begin{equation}
\langle i',m_z;j',-m_z|i,m_z;j,-m_z\rangle = \delta_{ii'}\delta_{jj'} \, ,
\end{equation}
\begin{equation}
\langle  i',m_z;j',-m_z|i,-m_z;j,m_z\rangle = \delta_{ii'}\delta_{jj'} \, .
\end{equation}
\end{subequations}

\noindent It means that within each set with a given $|m_z|=1$ or $2$, the states are orthogonal. However, there is no orthogonality between the states with $|m_z|=1$ and $|m_z|=2$:

\begin{eqnarray}\label{overlap12}
\nonumber
&&\langle i',+2;j',-2|i,+1;j,-1\rangle\\
\nonumber
&&=\sum_{\{{\bm k}\}} \tilde\varphi^*_{i'}(\bm k_{e_2}, \bm k_{h_1}) \tilde\varphi^*_{j'}(\bm k_{e_1}, \bm k_{h_2}) \tilde\varphi_{i}(\bm k_{e_1}, \bm k_{h_1}) \tilde\varphi_{j}(\bm k_{e_2}, \bm k_{h_2})\\
&&=\lambda\left(
\begin{array}{cc}
i' & j \\
j' & i \\
\end{array}\right),
\end{eqnarray}

\noindent where, for ease of notations, $\{{\bm k}\} \equiv {\bm k_{e_1}, \bm k_{e_2},\bm k_{h_1}, \bm k_{h_2}}$. All other needed overlaps follow directly from Eq. \eqref{overlap12}.

\subsection{Scattering via dark states}

We consider two channels for the scattering of excitons with opposite spins. The first scattering process appears in the Born approximation and is described by the matrix element

\begin{equation}
 \label{direct}
\mathcal M_{i \to f}^{(1)} = \langle{\rm vac}| B_{1',+1} B_{2',-1}   H  B_{1,+1}^\dag B_{2,-1}^\dag |{\rm vac}\rangle.
\end{equation}

We remind that the initial and final states $|i\rangle = B_{1,+1}^\dag B_{2,-1}^\dag |{\rm vac}\rangle$ and $|f\rangle = B_{1',+1}^\dag B_{2',-1}^\dag |{\rm vac}\rangle$ are orthogonal and hence an additional contribution due to non-orthogonality is absent in Eq. \eqref{direct}. This matrix element can be calculated immediately, see Ref.~\cite{COM08}, Eqs. (8.25), (8.27) and (8.30):

\begin{equation}
 \label{direct:val}
\mathcal M_{i \to f}^{(1)} = \xi
\left(\begin{array}{cc}
2' & 2 \\
1' & 1 \\
\end{array}\right).
\end{equation}

Equation \eqref{direct:val} is nothing but a \emph{direct} Coulomb interaction between the excitons. In the case of cavity polaritons this contribution can be extremely small for polaritons as $\xi(\bm Q) \propto (Q a_B)^{3}$, where $\bm Q$ is the transferred wavevector. In the case of cavity polaritons the matrix element of Eq. \eqref{direct:val} should be multipled by the product of Hopfield coefficients of initial and final states.

Another possibility to have an interaction between excitons with opposite spins is to consider the scattering via intermediate dark states \footnote{The light-matter coupling term in Eq.~\eqref{Hlmc} does not yield additional contributions to the scattering of polaritons with opposite spins.} with $|m_z|=2$. In the case of excitons the splitting between the dark and bright states (determined mostly by the short-range electron-hole exchange) is usually small as compared with the typical energy of excitons; therefore such a scattering populates the dark states. This is not the case in microcavities: the splitting between dark and bright states can be quite large (of the order of the Rabi splitting) and hence the dark states can play the role of intermediate, virtual states in the polariton-polariton scattering process. The related matrix element is weakly dependent on the exciton wavevectors, but it appears at the next order of perturbation theory.

We carry out the calculation of the scattering rate between the polaritons with opposite spins assuming that the splitting between polariton states and dark excitons states, $\Delta$, exceeds by far the kinetic energy of polaritons, $E_k^{P}$. This condition is fulfilled in quantum microcavities under the resonant excitation of the lower polariton branch below or around the inflection point.

The matrix elements of the interaction induced transitions from bright to dark states are needed. This process takes place due to the exchange of electrons (or holes) between excitons which results in the transfer of excitons from bright to the dark states. The Hamiltonian matrix element can be calculated using the fermionic approach as follows:

\begin{eqnarray}
\label{1to2}
&&\langle{\rm vac}| B_{1',+2} B_{2',-2}   H_{\rm exc}  B_{1,+1}^\dag B_{2,-1}^\dag |{\rm vac}\rangle =
\nonumber\\
&&(E_1+E_2) \langle{\rm vac}| B_{1',+2} B_{2',-2}  B_{1,+1}^\dag B_{2,-1}^\dag |{\rm vac}\rangle
\nonumber\\
&+&\sum_{nm} \xi
\left(
\begin{array}{cc}
 n & 2\\
 m & 1\\
\end{array}
\right)
\langle{\rm vac}| B_{1',+2} B_{2',-2}
B_{m,+1}^\dag B_{n,-1}^\dag |{\rm vac}\rangle
\nonumber\\
&=&(E_1+E_2)\lambda
\left(
\begin{array}{cc}
 1' & 2\\
 2' & 1\\
\end{array}
\right) +  \xi^{in}
\left(
\begin{array}{cc}
 1' & 2\\
 2' & 1\\
\end{array}
\right).
\end{eqnarray}

\noindent It is clear that such a process is solely determined by the exchange interaction.

In order to calculate the matrix element of the exciton-exciton scattering via dark states we need: (i) to orthogonalize the intermediate states with respect to the initial and final states, and (ii) take into account the indistinguishability of the states $|i,m_z;j,-m_z\rangle$ and $|j,-m_z;i,+m_z\rangle$. The first point can be easily achieved by the subtracting from the states of the pairs of excitons with $|m_z|=2$ the admixture of the states where $|m_z|=1$. The corresponding matrix element between the orthogonalized states differs from Eq. \eqref{1to2} by the absence of the first term (which is proportional to $E_1+E_2$)~\footnote{It is assumed that the direct interaction described by Eq. \eqref{direct} is negligibly small.}. The second point can be dealt with by using the proper symmetrization of the two-polariton wavefunction. In the case of opposite spin configurations one can introduce either symmetric ($\rm s$) or antisymmetric ($\rm a$) combinations as

\begin{subequations}
\begin{equation}\label{lst}
|i,j;|m_z|\rangle_{\rm s} = \frac{1}{\sqrt{2}}\left(|i,m_z;j,-m_z\rangle + |i,-m_z;j,m_z\rangle\right)
\end{equation}
\begin{equation}
|i,j;|m_z|\rangle_{\rm a} = \frac{1}{\sqrt{2}}\left(|i,m_z;j,-m_z\rangle - |i,-m_z;j,m_z\rangle\right)
\end{equation}
\end{subequations}

In the relevant wavevector range, the exciton-exciton scattering matrix elements weakly depend on the values of the wavevectors therefore the exciton-exciton interaction may be considered as short-range. In this case the interaction in the antisymmetric state is strongly suppressed as compared with the interaction in the symmetric state because the overlap of the wavefunctions of excitons in the former state is small.

In what follows let us consider the scattering of two excitons in a symmetric configuration, Eq.~\eqref{lst}. Let us denote $|i_1,i_2;1\rangle_{\rm s}$ the initial state, $|f_1,f_2;1\rangle_{\rm s}$ the final state. We denote as $|\widetilde{j_1,j_2};2 \rangle_{\rm s}$ the states of pairs of dark excitons in a symmetric configuration which are orthogonalized with respect to the initial and final states. The matrix element of the transition, calculated to the second order of perturbation theory, can be presented as~\cite{ll3_eng}

\begin{eqnarray}
\label{2ord}
\mathcal M_{i \to f}^{(2)} & = &{\sum_{j_1,j_2}}' \left(E_{i_1} + E_{i_2} - E_{j_1} - E_{j_2}\right)^{-1}
\\
&\times& \langle f_1, f_2;1| H| \widetilde{j_1,j_2};2\rangle \langle \widetilde{j_1,j_2};| H |i_1,i_2;1\rangle.
\nonumber
\end{eqnarray}

\noindent Here the summation is carried out over the different states (the states $|j_1,j_2;2\rangle_{\rm s}$ and $|j_2,j_1;2\rangle_{\rm s}$ are actually the same), and the subscript $\rm s$ is omitted. Using Eq. \eqref{1to2} this expression can be recast as

\begin{eqnarray}
\label{2ord1}
\mathcal M_{i \to f}^{(2)} & = & {\sum_{j_1,j_2}}' ({E_{i_1} + E_{i_2} - E_{j_1} - E_{j_2}})^{-1}
\nonumber\\
&\times&
\left[
\xi^{out}
\left(
\begin{array}{cc}
f_2 & j_2\\
f_1 & j_1\\
\end{array}
\right) + \xi^{in}
\left(
\begin{array}{cc}
f_1 & j_2\\
f_2 & j_1\\
\end{array}
\right)
\right]
\nonumber\\
&\times&
\left[\xi^{out}
\left(
\begin{array}{cc}
 j_2 & i_2\\
 j_1 & i_1\\
\end{array}
\right) + \xi^{in}
\left(
\begin{array}{cc}
 j_1 & i_2\\
 j_2 & i_1\\
\end{array}
\right) \right],
\end{eqnarray}

\noindent where the quantity $\xi^{out}$ is defined as

\[
\xi^{out}
\left(
\begin{array}{cc}
 n & j\\
 m & i\\
\end{array}
\right)
=
\left[\xi^{in}
\left(
\begin{array}{cc}
 j & n\\
 i & m\\
\end{array}
\right)\right]^*,
\]

\noindent and represent the Coulomb exchange scattering between the \emph{out} excitons~\cite{COM07,COM08}. It is worth noting that the counterpart of Eq.~\eqref{2ord1} for the antisymmetric wavefunctions differs by the signs in the square brackets only. We mention that since for free excitons in quantum wells, dark-bright splitting is usually very small as compared to the exciton kinetic energy, Eq.~\eqref{2ord1} cannot be used and the proper calculation of the matrix element to all orders of perturbation theory is needed. On the other hand, in microcavities the splitting between dark excitons and polaritons may be huge (of the order of the Rabi splitting), which is why the extension of Eq.~\eqref{2ord1} to the case of the cavity polaritons is perfectly applicable.

This extension is straightforward: the energies for the initial states in the denominator should be polariton energies, and energies of intermediate states are not affected by the light-matter coupling simply because these states are dark. Moreover the product of Hopfield factors $\beta_{i_1}\beta_{i_2}\beta_{f_1}^*\beta_{f_2}^*$ should be introduced in Eq.~\eqref{2ord1}. Finally, the result for polaritons reads:

\begin{eqnarray}
\label{2ord1p}
\mathcal M_{i \to f}^{(2,{\rm pol})} & = & \beta_{i_1}\beta_{i_2}\beta_{f_1}^*\beta_{f_2}^*{\sum_{j_1,j_2}}' ({E_{i_1}^P + E_{i_2}^P - E_{j_1}^P - E_{j_2}^P})^{-1}
\nonumber\\
&\times&
\left[
\xi^{out}
\left(
\begin{array}{cc}
f_2 & j_2\\
f_1 & j_1\\
\end{array}
\right) + \xi^{in}
\left(
\begin{array}{cc}
f_1 & j_2\\
f_2 & j_1\\
\end{array}
\right)
\right]
\nonumber\\
&\times&
\left[\xi^{out}
\left(
\begin{array}{cc}
 j_2 & i_2\\
 j_1 & i_1\\
\end{array}
\right) + \xi^{in}
\left(
\begin{array}{cc}
 j_1 & i_2\\
 j_2 & i_1\\
\end{array}
\right) \right].
\end{eqnarray}

Equation~\eqref{2ord1p} is the central result of this section. It is a microscopic derivation of the polariton-polariton scattering effective matrix element in the anti-parallel spin configuration, introduced in a number of works, see Refs.~\cite{rot,inv}.

It was demonstrated in Ref.~\cite{inv} that the negative relative sign of matrix elements for polariton-polariton scattering in the parallel and anti-parallel spin configurations can lead to an inversion of the linear polarization in microcavities in the non-linear regime. Provided that the main contribution to the polariton-polariton scattering is determined by the exciton-exciton interaction which, in turn, is determined by the repulsive exchange term $\xi^{in}$ (see Sec.~\ref{sec:est}), the relative sign of the effective matrix elements is given by the sign of the energy denominator: $({E_{i_1}^P + E_{i_2}^P - E_{j_1}^P - E_{j_2}^P})$. Obviously, if the dark states lie above the bright ones (i.e. we consider the scattering of polaritons of the \emph{lower} branch) the denominator is negative which corresponds to an effective attraction of polaritons and to the inversion of linear polarization. On the \emph{upper} branch the dark states are lower in energy, therefore the denominator is positive, the polariton-polariton interaction is repulsive and the inversion of linear polarization is not observed.

\section{Estimations of scattering matrix elements}\label{sec:est}

Knowledge, even approximate, of the scattering matrix elements in contexts relevant to experiments is instructive. In this section we provide and discuss simple estimations obtained for both parallel and opposite spin configurations. Detail of the evaluations of the integrals $\lambda$ and $\rho_{\bm \varkappa}(i,j;k)$, respectively defined in Eqs.~\eqref{lambda} and \eqref{rho}, is given in Appendix A.

\subsection{Estimation of the scattering matrix elements in the parallel spin configuration}

\subsubsection{Direct scattering}

As in Ref.~\cite{COM07}, we find that the direct term of the scattering matrix element, $\xi(\bm Q)$, as a function of transferred wavevector $\bm Q$, can be expressed as follows:

\begin{equation}\label{xi_dir}
\xi(\bm Q) = V_{\bm Q} \left[g(\gamma_e\bm Q) - g(\gamma_h \bm Q) \right]^2,
\end{equation}

\noindent where $\gamma_e = m_e/M_{\rm exc}$, $\gamma_h = m_h/M_{\rm exc}$, with $m_e$, $m_h$, $M_{\rm exc}= m_e + m_h$ being the electron, hole and exciton effective masses respectively; $g(\bm Q) = (1+a_B^2Q^2/4)^{-3/2}$ is the Fourier transform of $\left|\psi(\bm \rho)\right|^2$, and $V_{\bm Q}$ is the Fourier transform of the 2D Coulomb interaction potential: 

\begin{equation}\label{coul_vq}
V_{\bm Q} = \frac{2\pi e^2}{\kappa S Q},
\end{equation}

\noindent $\kappa$ being the background dielectric constant.

In the limit $Qa_B \ll 1$, Eq.~\eqref{xi_dir} reduces to

\begin{equation}
\label{xi_dir1}
\xi(\bm Q) = \frac{e^2 }{\kappa a_B} \frac{a_B^2}{S} \frac{9\pi }{32} (\gamma_e^2-\gamma_h^2)^2 (Q a_B)^3
\end{equation}

\subsubsection{Exchange scattering}

Denoting $\tilde{\varphi}(\bm k)$ the Fourier transform of the 2D exciton ground state wavefunction Eq.~\eqref{psi}:

\begin{equation}
\label{phik}
\tilde{\varphi}(\bm k) = \sqrt{\frac{2\pi}{S}} \frac{2a_B}{[1+ (k a_B)^2]^{3/2}},
\end{equation}

\noindent we find that the Coulomb exchange scattering between \emph{in} excitons, Eq.~\eqref{xi_in}, can be recast as

\begin{equation}\label{xi:in}
\xi^{in} = \xi^{in}_{ee} + \xi^{in}_{hh} + \xi^{in}_{eh} + \xi^{in}_{he},
\end{equation}

\noindent where

\begin{widetext}
\begin{subequations}
\begin{eqnarray}
\xi^{in}_{ee} = \xi^{in}_{hh} = \phantom{-}\sum_{\bm k,\bm q} V_{\bm q} \tilde{\varphi}{\left(\bm k + \frac{\bm P_{-} + \bm q}{2}\right)}\tilde{\varphi}{\left(\bm k - \frac{\bm P_{-} + \bm q}{2}\right)} \tilde{\varphi}{\left(\bm k + \frac{\bm P_{+} - \bm q}{2}\right)} \tilde{\varphi}{\left(\bm k - \frac{\bm P_{+} - \bm q}{2}\right)}, \label{ee}\\
\xi^{in}_{eh} = \xi^{in}_{he} = -\sum_{\bm k,\bm q} V_{\bm q} \tilde{\varphi}{\left(\bm k + \frac{\bm P_{-} + \bm q}{2}\right)}\tilde{\varphi}{\left(\bm k - \frac{\bm P_{-} + \bm q}{2}\right)} \tilde{\varphi}{\left(\bm k + \frac{\bm P_{+} + \bm q}{2}\right)} \tilde{\varphi}{\left(\bm k - \frac{\bm P_{+} - \bm q}{2}\right)}, \label{eh}
\end{eqnarray}
\end{subequations}
\end{widetext}

\noindent with
\[
\bm P_{\pm} = 2\alpha_h \bm p + (\alpha_h \pm \alpha_e) \bm Q,
\]

\noindent and $\bm p$ is the wavevector characterizing the relative motion of excitons.

The general expression of these four contributions to the scattering matrix element $\xi^{in}$ are quite complicated. However, taking into account that the wavevectors of the exciton-polariton center of mass motion are very small compared to the inverse Bohr radius, we can simply assume that the moduli of the vectors $\bm p$, $\bm Q$ (and, therefore, $\bm P_{\pm}$) are negligible. It follows that

\begin{subequations}
\begin{eqnarray}
\xi^{in}_{ee} = \xi^{in}_{hh} = \sum_{\bm k,\bm q} V_{\bm q} \tilde{\varphi}^2{\left(\bm k + \frac{\bm q}{2}\right)}\tilde{\varphi}^2{\left(\bm k - \frac{\bm q}{2}\right)}, \label{ee1}\\
\xi^{in}_{eh} = \xi^{in}_{he} = -\sum_{\bm k,\bm q} V_{\bm q} \tilde{\varphi}^3{\left(\bm k + \frac{\bm q}{2}\right)}\tilde{\varphi}{\left(\bm k - \frac{\bm q}{2}\right)}. \label{eh1}
\end{eqnarray}
\end{subequations}

Under the assumption that excitons are strictly two-dimensional so that Coulomb interaction is described by the potential Eq.~\eqref{coul_vq} we obtain

\begin{subequations}
\begin{eqnarray}\label{xi_res}
 \xi^{in}_{ee} + \xi^{in}_{hh} &\approx & \phantom{-}19\frac{e^2}{\kappa a_B} \frac{a_B^2}{S}\\
 \xi^{in}_{eh} + \xi^{in}_{he} &\approx &-25\frac{e^2}{\kappa a_B} \frac{a_B^2}{S},\label{xi_res_b}
\end{eqnarray}
\end{subequations}

\noindent in agreement with Ref.~\cite{TAS99}. It is also instructive to compare the amplitudes of direct scattering and exchange scattering terms:

\begin{equation}
\frac{\xi(q)}{\xi^{in}} \approx 0.15(\gamma_e^2-\gamma_h^2)^2 (q a_B)^3 \sim 1.5\times 10^{-4},
\end{equation}

\noindent which shows that the exchange mechanism dominates the scattering process. To make the latter estimation we used the following values: $Q \sim 10^{5}$~cm$^{-1}$ and $a_B \sim 10^{-6}$~cm. Note that both direct and exchange scattering processes result in repulsive interactions of polaritons.

\subsection{Estimation of the scattering matrix elements in the opposite spin configuration}

The effective matrix element of the polariton-polariton interaction with opposite spins is the most difficult quantity to estimate. The problem stems from the fact that the dark states are involved and, as a consequence, the transferred momentum in the process of virtual transition to the dark states can be arbitrary. The exact evaluation of the effective scattering matrix elements will be a subject of a future work. Here we present an order-of-magnitude estmation of the considered effect. 

The initial and final states of an exciton are those with small momenta $K, K' \ll 1/a_B$. Therefore these momenta can be neglected and one can consider the scattering of two excitons from the ground state to the dark states with the momenta $\bm P$ and $-\bm P$. Therefore, Eq. \eqref{2ord1p} can be rewritten as

\begin{equation}
\label{est1}
\mathcal M_{i \to f}^{(2,{\rm pol})} = h_{i\to f}\sum_{\bm P} \frac{|U(\bm P)|^2}{-\Delta - \hbar^2 P^2/M_X}.
\end{equation}
Here $h_{i\to f}$ is a product of Hopfield factors for initial and final states, cf. Eq.~\eqref{2ord1p}, $\Delta$ is the splitting between dark and bright states, and
 
\[
 U(\bm P) = \left[\xi^{out}
\left(
\begin{array}{cc}
 \phantom{-}\bm P & 0\\
 -\bm P & 0\\
\end{array}
\right) + \xi^{in}
\left(
\begin{array}{cc}
 \phantom{-}\bm P & 0\\
 -\bm P & 0\\
\end{array}
\right) \right].
\]

\noindent The summation over $\bm P$ can be transformed into an integral as

\begin{equation}
\label{est2}
\mathcal M_{i \to f}^{(2)} = h_{i\to f}\frac{S}{2\pi} \frac{M_X}{2\hbar^2} \int \mathrm d \varepsilon \frac{|U(\varepsilon)|^2}{-\Delta - \varepsilon}
\end{equation}

To make an estimation we assume the following model for $U(\varepsilon)$: $U(\varepsilon) = U_0$ for $\varepsilon<\varepsilon_0$ and $0$ otherwise. An integration in Eq.~\eqref{est2} yields

\begin{equation}
\label{est3}
\mathcal M_{i \to f}^{(2)} = -h_{i\to f}\frac{S}{2\pi} \frac{M_X}{2\hbar^2} U_0^2 \ln{\left( \frac{\Delta + \varepsilon_0}{\Delta}\right)}.
\end{equation}

\noindent We set $U_0 = e^2/(\kappa a_B)(a_B^2/S)$ (see Eqs.~\eqref{xi_res} and \eqref{xi_res_b}), and $\varepsilon_0 = \hbar^2/(M_X a_B^2)$ (because the wavevector cut-off is at $P\sim 1/a_B$ and, hence, the cut-off energy is $\hbar^2K^2/M_X$). Thus, one obtains
\begin{equation}
\label{est_fin}
\mathcal M_{i \to f}^{(2)} \sim -h_{i\to f}\left(\frac{e^2}{\kappa a_B}\right)^2 \frac{a_B^2}{S} \frac{M_X a_B^2}{\hbar^2} \ln{\left( 1 + \frac{\hbar^2}{M_X a_B^2\Delta}\right)}.
\end{equation}

\noindent The numerical factor is omitted here. 

Depending on the relation between $\varepsilon = \hbar^2/(M_X a_B^2)$ and $\Delta$ there are two limits:

\begin{enumerate}
\item if $\varepsilon = \frac{\hbar^2}{2 M_X a_B^2} \ll \Delta$,
we obtain

\begin{equation}
\label{est_fin:1}
\mathcal M_{i \to f}^{(2)} \sim -h_{i\to f}\frac{1}{\Delta}\left(\frac{e^2}{\kappa a_B}\right)^2 \frac{a_B^2}{S}.
\end{equation}

\item if $\varepsilon = \frac{\hbar^2}{2 M_X a_B^2} \gg \Delta$,
we obtain

\begin{equation}
\mathcal M_{i \to f}^{(2)} \sim -\left(\frac{e^2}{\kappa a_B}\right)^2 \frac{a_B^2}{S} \frac{M_X a_B^2}{\hbar^2} \ln{\left(\frac{\hbar^2}{M_X a_B^2\Delta}\right)}.
\end{equation}

\end{enumerate}

So far we have assumed that the light-matter coupling is relatively strong, so that the dark-bright states splitting $\Delta$, being equal to half the Rabi splitting, exceeds by far the matrix elements of polariton-polariton interaction. In this case we can limit the calculation of the scattering amplitude in the opposite spins configuration to the second order of perturbation theory. The situation is different if the light-matter coupling strength is small.

In order to gain a qualitative understanding of the situation with small values of $\Delta$, one can neglect the light-matter interaction entirely and consider the interaction of bare excitons. The exciton-exciton scattering via excited states is discussed in Ref.~\cite{inoue00}, where it is demonstrated that an interaction of polaritons with opposite spins can result from the scattering via such states as $2s, 1p, \ldots$ An estimation given in \cite{inoue00} has the same dimensional parameters as Eq. \eqref{est_fin}.

Moreover, the condition $K a_B \ll 1$ shows that the scattering takes place in the so-called low energy region. Thus, the scattering rates can be strongly enhanced. Under the assumption that the Rabi splitting is smaller than the bi-exciton (bi-polariton) binding energy, one may use the results of a general theory of two-dimensional scattering and write~\cite{TAK02,averbuch86,verhaar,shumway06}

\begin{equation}\label{general_sc2d}
\mathcal M_{i\to f} = \frac{4\pi \hbar^2}{M_X S} \ln{\left( -\frac{\varepsilon}{E_0} \right)}.
\end{equation}

\noindent Here $\varepsilon$ is the energy of the relative motion of a pair of excitons (or polaritons), $E_0>0$ is the bi-exciton (bi-polariton) binding energy. Note that scattering via bi-exciton or bi-polariton states \cite{IVA04} may be even more efficient than the scattering via dark states.

\section{Conclusion}
The scattering of excitons and exciton-polaritons in two-dimensional semiconductor microcavities has been considered here in the framework of the fermionic commutation technique Ref.~\cite{COM07}. Our results can be summarized as follows:
\begin{enumerate}
\item The basis of two-excitonic states is overcomplete which requires an orthogonalization of initial and final wavefunctions for the calculation of the matrix elements and scattering rates.
\item For a fixed finite number of inital and final states the scattering rates calculated within the fermionic approach and bozonization scheme are equivalent.
\item In the low-density regime the bosonic enhancement of exciton-exciton scattering is reproduced in the fermionic approach.
\item An additional contribution to polariton-polariton scattering rate which arises from the non-linearity in light-matter interaction is derived.
\item The scattering matrix elements for exciton-polaritons with anti-parallel spin configurations are calculated.
\end{enumerate}

\begin{acknowledgments}
We acknowledge the support of the EPSRC and the Agence Nationale de la Recherche. M. M. G. was partially supported by ``Dynasty'' Foundation -- ICFPM, President grant for young scientists programmes of RAS and RFBR.
\end{acknowledgments}

{\appendix

\section{Evaluation of $\rho$ and $\lambda$ overlap integrals}

To evaluate the integral $\rho_{\bm \varkappa}(i,j;k)$ as defined in Eq.~\eqref{rho}, we assume that the exciton wavefunction takes the following form:

\begin{equation}
 \varphi_a(\bm r_e, \bm r_h) = \frac{1}{\sqrt{S}} e^{\mathrm i \bm \varkappa_a \cdot \bm R} \psi(\bm \rho),
\end{equation}

\noindent where $\bm \varkappa_a$ is the exciton center of mass wavevector, $\bm R = \gamma_e \bm r_e + \gamma_h \bm r_h$ is the position of the center of mass and $\bm \rho = \bm r_e - \bm r_h$ is the electron-hole relative coordinate. Here $S$ is the normalization area, $\psi(\bm \rho)$ is the relative motion wavefunction. Substituting $\bm \rho_1 = \bm r_e - \bm r$ and $\bm \rho_2 = \bm r - \bm r_h$ and integrating over $\bm r$ we obtain

\begin{widetext}
\begin{multline}\label{rho2}
 \rho_{\bm \varkappa}(i,j;k) = \frac{1}{\sqrt{S}} \delta_{\bm \varkappa_i + \bm \varkappa_j,\bm \varkappa_0 + \bm \varkappa} \int \mathrm d \bm \rho_1 \int \mathrm d \bm \rho_2  \psi(\bm \rho_1) \psi(\bm \rho_2) \psi(\bm \rho_1 + \bm \rho_2) \times\\
 \exp{\left[\mathrm i \bm \rho_1 \cdot (- \gamma_e \bm \varkappa_i - \gamma_e \bm \varkappa_j + \gamma_e \bm \varkappa_0) \right]} \exp{\left[-\mathrm i \bm \rho_2 \cdot (- \gamma_h \bm \varkappa_i - \gamma_h \bm \varkappa_j + \gamma_h \bm \varkappa_0) \right]}.
\end{multline}
\end{widetext}

The characteristic length of the $\psi$ function variation is the excitonic Bohr radius, while typical values of the wavenumbers $\varkappa_{i,j,0} \ll a_B^{-1}$ because the polariton wavevectors are determined by the photon wavevectors. Therefore, in the integrations over $\bm \rho_1$ and $\bm \rho_2$ one may omit the exponential terms, which yields:

\begin{eqnarray}\label{rho:fin}
\rho_{\bm \varkappa}(i,j;k) & = & \frac{\delta_{\bm \varkappa_i + \bm \varkappa_j,\bm \varkappa_0 + \bm \varkappa} }{\sqrt{S}}
\\
&& \times \int \mathrm d \bm \rho_1 \int \mathrm d \bm \rho_2  \psi(\bm \rho_1) \psi(\bm \rho_2) \psi(\bm \rho_1 + \bm \rho_2).
\nonumber
\end{eqnarray}

Assuming the following form for the wavefunction of the relative electron-hole motion in the exciton 

\begin{equation}
 \label{psi}
\psi(\bm \rho) = \sqrt{\frac{2}{\pi a_B^2}}~\exp\left(-\rho/a_B\right),
\end{equation}

\noindent and introducing the dimensionless quantity

\begin{multline}
 \mathcal I = \int \mathrm d \rho_1 \mathrm d \rho_2 \int \mathrm d \theta \rho_1 \rho_2 \times \\ e^{-\rho_1} e^{-\rho_2} \exp{\left[-\sqrt{\rho_1^2 + \rho_2^2 +2 \rho_1 \rho_2 \cos{\theta}}\right]} \approx 0.896,
\end{multline}

\noindent we obtain

\begin{equation}\label{final}
 \rho_{\bm \varkappa}(i,j;k) =  \frac{4\Omega_R \mathcal I a_B^2}{SW}~\delta_{\bm \varkappa_i + \bm \varkappa_j,\bm \varkappa_0 + \bm \varkappa},
\end{equation}

\noindent where $W$ is defined in Eq.~\eqref{Hlmc}.

In the same manner the $\lambda$ overlap integral

\begin{widetext}
\begin{equation}\label{eq22}
\lambda = \int {\rm d}{\bf r}_{e_1}{\rm d}{\bf r}_{e_2}{\rm d}{\bf r}_{h_1}{\rm d}{\bf r}_{h_2}\varphi_m^*({\bf r}_{e_1},{\bf r}_{h_2})\varphi_n^*({\bf r}_{e_2},{\bf r}_{h_1})\varphi_i({\bf r}_{e_1},{\bf r}_{h_1})\varphi_j({\bf r}_{e_2},{\bf r}_{h_2}).
\end{equation}

\noindent reduces to the following expression
 
\begin{equation}\label{lambda:fin}
 \frac{1}{S}~\delta_{\bm \varkappa_i + \bm \varkappa_j,\bm \varkappa_m + \bm \varkappa_n} \int \mathrm d \bm \rho_1 \int \mathrm d \bm \rho_2 \int \mathrm d \bm r  \psi(\bm \rho_1 + \bm r) \psi(\bm \rho_1) \psi(\bm \rho_2  - \bm r) \psi(\bm \rho_2),
\end{equation}
\end{widetext}

\noindent which can be further simplified:

\begin{equation}\label{lambda:fin:1}
\lambda = \frac{8a_B^2}{\pi S} \delta_{\bm \varkappa_i + \bm \varkappa_j,\bm \varkappa_m + \bm \varkappa_n} \mathcal J,
\end{equation}

\noindent where $\mathcal J$ is given by

\begin{eqnarray}
\mathcal J & = &\int \mathrm d \rho_1 \mathrm d \rho_2  \int \mathrm d r \int \mathrm d \theta_1 \int \mathrm d \theta_2\, \rho_1 \rho_2 r \, e^{-\rho_1} e^{-\rho_2}
\nonumber\\
&&\times\exp{\left[-\sqrt{\rho_1^2 + r^2 -2 \rho_1 r \cos{\theta_1}}\right]}
\\
&&\times\exp{\left[-\sqrt{\rho_2^2 + r^2 +2 \rho_2 r \cos{\theta_2}}\right]}
\nonumber
\end{eqnarray}

\noindent and is numerically evaluated: $\mathcal J \approx 3.95$.

\section{Calculation of the scalar product involving three particle states}\label{app:scal}

Using the commutation rules outlined in Eqs. (6.32)--(6.35) of \cite{COM08}, we can exchange the order of $B_2$ and $(B_0^\dag)^2$ in Eq.~\eqref{bos:scalar} and, afterwards, swap $B_2$ and $B_1^\dag$. Finally, there are two types of terms that do not vanish:

\begin{equation}\label{bos:scalar:1}
\sum_n 2 \langle{\rm vac}|B_1^2B_0^\dag B_n^\dag|{\rm vac}\rangle \left[
\lambda\left(
\begin{array}{cc}
n & 0 \\
2 & 1 \\
\end{array}\right)
+
\lambda\left(
\begin{array}{cc}
2 & 0 \\
n & 1 \\
\end{array}\right)
\right],
\end{equation}

\noindent and

\begin{equation}\label{bos:scalar:2}
-\sum_n 2 \langle{\rm vac}|B_1^2 B_n^\dag B_1^\dag |{\rm vac}\rangle
\lambda\left(
\begin{array}{cc}
n & 0 \\
2 & 0 \\
\end{array}\right).
\end{equation}

\noindent There is a significant difference between the magnitudes of the terms given in Eqs. \eqref{bos:scalar:1} and \eqref{bos:scalar:2}. The contribution \eqref{bos:scalar:1} is small as compared to \eqref{bos:scalar:2} because the latter is determined by the term with $n=1$, where $\langle{\rm vac}|B_1^2 B_1^\dag B_1^\dag |{\rm vac}\rangle \approx 2$, while the magnitudes of all the terms in \eqref{bos:scalar:1} are further reduced because of the factor $\langle{\rm vac}|B_1^2B_0^\dag B_n^\dag|{\rm vac}\rangle \propto \nu$. Retaining only term with $n=1$ in Eq.~ \eqref{bos:scalar:2} we obtain the right hand side of Eq.~\eqref{bos:scalar}.

\section{Calculation of the matrix element of Hamiltonian involving three-particle states}\label{app:ham}

Performing the commutation of the operators $ H_{\rm exc}$ and $(B_0^\dag)^2$ in Eq.~\eqref{bos:h0}, we obtain four terms
\begin{subequations}\label{bos:scdev}
\begin{eqnarray}
&\langle{\rm vac}|B_1^2B_2  H_{\rm exc} (B_0^\dag)^2B_1^\dag|{\rm vac}\rangle  \nonumber\\
=& \langle{\rm vac}|B_1^2B_2 (B_0^\dag)^2 H_{\rm exc} B_1^\dag|{\rm vac}\rangle\\
+& 2E_0 \langle{\rm vac}|B_1^2B_2 (B_0^\dag)^2 B_1^\dag|{\rm vac}\rangle\\
+& 2 \langle{\rm vac}|B_1^2B_2 B_0^\dag V_0^\dag B_1^\dag|{\rm vac}\rangle  \label{bos:c0}\\
+& \sum\limits_{nm}\xi\left(
\begin{array}{cc}
n & 0 \\
m & 0 \\
\end{array}\right)
\langle{\rm vac}|B_1^2B_2 B_m^\dag B_n^\dag B_1^\dag|{\rm vac}\rangle. \label{bos:d}
\end{eqnarray}
\end{subequations}
The first two terms on the right hand side of the equation above, which we denote $a$ and $b$ can be evaluated easily:
\begin{subequations}
\begin{equation}\label{bos:a}
a = E_1  \langle f|i\rangle.
\end{equation}
\begin{equation}\label{bos:b}
b = 2E_0  \langle f|i\rangle.
\end{equation}
Term Eq.~\eqref{bos:c0} can be recast in a similar to the term Eq.~\eqref{bos:d} form:
\begin{equation}\label{bos:c}
c = 2\sum\limits_{nm}
\xi\left(
\begin{array}{cc}
n & 0 \\
m & 0 \\
\end{array}\right)
\langle{\rm vac}|B_1^2B_2 B_0^\dag B_m^\dag B_n^\dag|{\rm vac}\rangle,
\end{equation}
\end{subequations}
and the term Eq.~\eqref{bos:d} can be recast as
\begin{subequations}
\begin{eqnarray}
d 
=& \langle{\rm vac}|B_n B_m (B_1^\dag)^2B_1 B_2^\dag |{\rm vac}\rangle^* \label{d:zero}\\
+& 2\langle{\rm vac}|B_n B_m B_1^\dag (1 - D_{11}) B_2^\dag |{\rm vac}\rangle^* \label{d:lead}\\
-& 2\sum\limits_k
\lambda^*\left(
\begin{array}{cc}
k & 1 \\
1 & 1 \\
\end{array}\right)
\langle{\rm vac}|B_n B_m B_k^\dag B_2^\dag |{\rm vac}\rangle^*\label{d:other},
\end{eqnarray}
\end{subequations}

where $D_{11}$ is the deviation-from-boson operator introduced in Ref.~\cite{COM07}. The leading order contribution ($\propto\nu$) is given by the first summand in Eq.~\eqref{d:lead}. It yields:
\begin{equation}\label{bos:d21:1}
d = 4
\xi\left(
\begin{array}{cc}
2 & 0 \\
1 & 0 \\
\end{array}
\right) - 
4\xi^{in}\left(
\begin{array}{cc}
2 & 0 \\
1 & 0 \\
\end{array}\right).
\end{equation} 
One can check that Eq. \eqref{d:zero} gives a zero contribution, and Eqs.~\eqref{d:other} and \eqref{bos:c} yield similar expressions but with extra $\lambda$ factors, which give corrections of the order of $\nu^2$ to the effective scattering rate.

\section{Matrix elements of exciton-photon interaction Hamiltonian}\label{app:d}

To calculate the contribution of absorption nonlinearity to the polariton-polariton scattering rates, knowledge of the overlap between the polariton state $P^\dag B_0^{\dag}|{\rm vac}\rangle$ and the two-exciton state $B_1^\dag B_2^{\dag}|{\rm vac}\rangle$, is useful:

\begin{widetext}
\begin{equation}\label{1to2trans}
\langle{\rm vac}| B_1 B_2 P^\dag B_0^{\dag}|{\rm vac}\rangle \propto \sum_{\bm k_1^e, \bm k_1^h} \sum_{\bm k_2^e, \bm k_2^h} \sum_{\bm k_0^e, \bm k_0^h} \sum_{\bm \varkappa_e, \bm \varkappa_h} \delta_{\bm \varkappa_e, - \bm \varkappa_h + \bm \varkappa} \int \mathrm d\bm r_{e_1}\mathrm d\bm r_{h_1} \int \mathrm d\bm r_{e_2}\mathrm d\bm r_{h_2} \int \mathrm d\bm r_{e_0}\mathrm d\bm r_{h_0}
\end{equation}
\[
\exp{\left[\mathrm i \bm k_1^e \bm r_{e_1} + \mathrm i \bm k_1^h \bm r_{h_1}
+ \mathrm i \bm k_2^e \bm r_{e_2} + \mathrm i \bm k_2^h \bm r_{h_2}
-\mathrm i \bm k_0^e \bm r_{e_0} -\mathrm i \bm k_0^h \bm r_{h_0}\right]}
\varphi_1^*(\bm r_{e_1}, \bm r_{h_1}) \varphi_2^*(\bm r_{e_2}, \bm r_{h_2})
\varphi_0(\bm r_{e_0}, \bm r_{h_0})\times
\]
\[
\langle{\rm vac}|a_{\bm k_1^e} a_{\bm k_2^e} a_{\bm \varkappa_e}^\dag a_{\bm k_0^e}^\dag |{\rm vac}\rangle
\langle{\rm vac}|b_{\bm k_1^h} b_{\bm k_2^h} b_{\bm \varkappa_h}^\dag b_{\bm k_0^h}^\dag |{\rm vac}\rangle =
\]
\[
\int \mathrm d \bm r \mathrm d \bm r_e \mathrm d \bm r_h \varphi_2^*(\bm r, \bm r) { e^{\mathrm i \bm \varkappa \bm r}} \varphi_1^*(\bm r_e, \bm r_h) \varphi_0 (\bm r_e, \bm r_h) +
\int \mathrm d \bm r \mathrm d \bm r_e \mathrm d \bm r_h \varphi_1^*(\bm r, \bm r) {e^{\mathrm i \bm \varkappa \bm r}}  \varphi_2^*(\bm r_e, \bm r_h) \varphi_0 (\bm r_e, \bm r_h)
\]
\[
 - 
\int \mathrm d \bm r \mathrm d \bm r_e \mathrm d \bm r_h  \varphi_1^*(\bm r_e, \bm r) \varphi_2^*(\bm r, \bm r_h) { e^{\mathrm i \bm \varkappa \bm r}} \varphi_0 (\bm r_e, \bm r_h)
-
\int \mathrm d \bm r \mathrm d \bm r_e \mathrm d \bm r_h  \varphi_2^*(\bm r_e, \bm r) \varphi_1^*(\bm r, \bm r_h) {e^{\mathrm i \bm \varkappa \bm r}} \varphi_0 (\bm r_e, \bm r_h),
\]

\noindent The above result was found using the following identity:

\begin{equation}\label{use}
\langle{\rm vac}|a_{\bm k_1^e} a_{\bm k_2^e} a_{\bm \varkappa_e}^\dag a_{\bm k_0^e}^\dag |{\rm vac}\rangle
\langle{\rm vac}|b_{\bm k_1^h} b_{\bm k_2^h} b_{\bm \varkappa_h}^\dag b_{\bm k_0^h}^\dag |{\rm vac}\rangle = 
\left(\delta_{\bm k_0^e, \bm k_1^e} \delta_{\bm \varkappa_e, \bm k_2^e} - 
\delta_{\bm k_0^e, \bm k_2^e} \delta_{\bm \varkappa_e, \bm k_1^e}
 \right) \times
\left(\delta_{\bm k_0^h, \bm k_1^h} \delta_{\bm \varkappa_h, \bm k_2^h} - 
\delta_{\bm k_0^h, \bm k_2^h} \delta_{\bm \varkappa_h, \bm k_1^h}
 \right).
\end{equation}

\end{widetext}
}


\begin{thebibliography}{99}
\expandafter\ifx\csname natexlab\endcsname\relax\def\natexlab#1{#1}\fi
\expandafter\ifx\csname bibnamefont\endcsname\relax
  \def\bibnamefont#1{#1}\fi
\expandafter\ifx\csname bibfnamefont\endcsname\relax
  \def\bibfnamefont#1{#1}\fi
\expandafter\ifx\csname citenamefont\endcsname\relax
  \def\citenamefont#1{#1}\fi
\expandafter\ifx\csname url\endcsname\relax
  \def\url#1{\texttt{#1}}\fi
\expandafter\ifx\csname urlprefix\endcsname\relax\def\urlprefix{URL }\fi
\providecommand{\bibinfo}[2]{#2}
\providecommand{\eprint}[2][]{\url{#2}}

\bibitem[{\citenamefont{Kavokin and Malpuech}(2003)}]{kavokin03b}
\bibinfo{author}{\bibfnamefont{A.}~\bibnamefont{Kavokin}} \bibnamefont{and}
\bibinfo{author}{\bibfnamefont{G.}~\bibnamefont{Malpuech}},
\emph{\bibinfo{title}{Cavity Polaritons}}, vol.~\bibinfo{volume}{32} of
\emph{\bibinfo{series}{Thin Films and Nanostructures}}
(\bibinfo{publisher}{Elsevier}, \bibinfo{year}{2003}); A. Kavokin, J. Baumberg, G. Malpuech, F. Laussy, \textit{Microcavities}, Clarendon Press Oxford (2006).
\bibitem{stim} P. G. Savvidis, J. J. Baumberg, R. M. Stevenson, M. S. Skolnick, D. M. Whittaker, and J. S. Roberts, Phys. Rev. Lett. \textbf{84}, 1547 (2000).
\bibitem{rot} D. N. Krizhanovskii, D. Sanvitto, I. A. Shelykh, M. M. Glazov, G. Malpuech, D.D. Solnyshkov, A. Kavokin, S. Ceccarelli, M. S. Skolnick, and J. S. Roberts, Phys. Rev. B {\bf 73}, 073303 (2006).
\bibitem{bistable} A. Baas, J.-Ph. Karr, M. Romanelli, A. Bramati, and E. Giacobino Phys. Rev. B {\bf 70}, 161307(R) (2004).
\bibitem{multistable} N. A. Gippius, I. A. Shelykh, D. D. Solnyshkov, S. S. Gavrilov, Y. G. Rubo, A. V. Kavokin, S. G. Tikhodeev, G. Malpuech, Phys. Rev. Lett. {\bf 98}, 236401 (2007).
\bibitem{malpuech} G. Malpuech, D. D. Solnyshkov, H. Ouerdane, M. M. Glazov, I. Shelykh, Phys. Rev. Lett. {\bf 98}, 206402 (2007).
\bibitem{yamamoto08}S. Utsunomiya, L Tian, G. Roumpos, C. W. Lai, N. Kumada, T. Fujisawa, M. Kuwata-Gonokami, A. Loffler, S Hofling, A. Forchel, Y. Yamamoto, Nat. Phys., {\bf 9}, 674 (2008). 
\bibitem{bose} J. Kasprzak, M. Richard, S. Kundermann, A. Baas, P. Jeambrun, J. M. J. Keeling, F. M. Marchetti, M. H. Szymanska, R. Andre, J. L. Staehli, V. Savona, P. B. Littlewood, B. Deveaud, and Le Si Dang Nature \textbf{443}, 409 (2006).
\bibitem{baumberg}	J. J. Baumberg, A.V. Kavokin, S. Christopoulos, A. J. D. Grundy, R. Butt\'e, G. Christmann, D. D. Solnyshkov, G. Malpuech, G. Baldassarri H\"oger von H\"ogersthal, E. Feltin, J.-F. Carlin, and N. Grandjean, Phys. Rev. Lett. {\bf 101}, 136409 (2008).
\bibitem{WOL93} L. Wolniewicz, J. Chem. Phys. {\bf 99}, 1851 (1993).
\bibitem{JAM00} M. J. Jamieson, A. Dalgarno, and L. Wolniewicz, Phys. Rev. A {\bf 61}, 042705 (2000).
\bibitem{JAM09} M. J. Jamieson, A. S.-C. Cheung, and H. Ouerdane, \textit{Dependence of the scattering length for hydrogen atoms on effective mass}, Euro. Phys. J. D, (to be published).
\bibitem{FEN87} Y. Feng and H. N. Spector, J. Phys. Chem. Solids {\bf 48}, 1191 (1987).
\bibitem{KOH97} T. S. Koh, Y. P. Feng, and H. N. Spector, Phys. Rev. B {\bf 55}, 9271 (1997).
\bibitem{HAN77} E. Hanamura and H. Haug, Phys. Rep. {\bf 33}, 209 (1977).
\bibitem{IVA98} A. L. Ivanov, H. Haug, and L. V. Keldysh, Phys. Rep. {\bf 296}, 237 (1998).
\bibitem{USU60} T. Usui, Progr. Theor. Phys. {\bf 23}, 787 (1960). 
\bibitem{CIU98} C. Ciuti, V. Savona, C. Piermarocchi, A. Quattropani and P. Schwendimann, Phys. Rev. B {\bf 58}, 7926 (1998).
\bibitem{TAS99} F. Tassone and Y. Yamamoto, Phys. Rev. B {\bf 59}, 10830 (1999).
\bibitem{inoue00} J. I. Inoue, T. Brandes, and A. Shimizu, Phys. Rev. B {\bf 61}, 2863 (2000).
\bibitem{BEN01} S. Ben-Tabou de-Leon and B. Laikhtman, Phys. Rev. B {\bf 63}, 125306 (2001).
\bibitem{OKU01} S. Okumura and T. Ogawa, Phys. Rev. B {\bf 65}, 035105 (2001).
\bibitem{AMA94} T. Amand, X. Marie, B. Baylac, B. Dareys, J. Barrau, M. Brousseau, R. Planel, and D. J. Dunstan, Phys. Lett. A {\bf 193}, 105 (1994).
\bibitem{KUW97} M. Kuwata-Gonokami, S. Inouye, H. Suzuura, M. Shirane, R. Shimano, T. Someya, and H. Sakaki, Phys. Rev. Lett. {\bf 79}, 1341 (1997).
\bibitem{AMA97} T. Amand, D. Robart, X. Marie, M. Brousseau, P. Le Jeune, and J. Barrau, Phys. Rev. B {\bf 55}, 9880 (1997).
\bibitem{LEJ98} P. Le Jeune, X. Marie, T. Amand, F. Romstad, F. Perez, J. Barrau, and M. Brousseau, Phys. Rev. B {\bf 58}, 4853 (1998).
\bibitem{BUT99} L. V. Butov, A. A. Shashkin, V. T. Dolgopolov, K. L. Campman, and A. C. Gossard, Phys. Rev. B {\bf 60}, 8753 (1999).
\bibitem{SCH08} C. Schindler and R. Zimmermann, Phys. Rev. B {\bf 78}, 045313 (2008).
\bibitem{LIN88} M. Lindberg and S. W. Koch, Phys. Rev. B {\bf 38}, 3342 (1988).
\bibitem{AXT94a} V. M. Axt and A. Stahl, Z. Phys. B {\bf 93}, 195 (1994).
\bibitem{AXT94b} V. M. Axt and A. Stahl, Z. Phys. B {\bf 93}, 205 (1994).
\bibitem{VIC95} K. Victor, V. M. Axt, and A. Stahl, Phys. Rev. B {\bf 51}, 14164 (1995).
\bibitem{OST95} Th. \"Ostreich, K. Sch\"onhammer, and L. J. Sham, Phys. Rev. Lett. {\bf 74}, 4698 (1995).
\bibitem{OST98} Th. \"Ostreich, K. Sch\"onhammer, and L. J. Sham, Phys. Rev. B {\bf 58}, 12920 (1998).
\bibitem{AXT01} V. M. Axt, S. R. Bolton, U. Neukirch, L. J. Sham, and D. S. Chemla, Phys. Rev. B {\bf 63}, 115303 (2001).
\bibitem{SAV96} S. Savasta and R. Girlanda, Phys. Rev. Lett. {\bf 77}, 4736 (1996).
\bibitem{SCH07} S. Schumacher, N. H. Kwong, and R. Binder, Phys. Rev. B {\bf 76}, 245324 (2007). 
\bibitem{POR08} S. Portolan, O. Di Stefano, S. Savasta, F. Rossi, and R. Girlanda Phys. Rev. B {\bf 77}, 195305 (2008).
\bibitem{TAK02} R. Takayama, N.H. Kwong, I. Rumyantsev, M. Kuwata-Gonokami, and R. Binder, Eur. Phys. J. B {\bf 25}, 445 (2002).
\bibitem{inv} K.V. Kavokin, P. Renucci, T. Amand, X. Marie, P. Senellart, J. Bloch, B. Sermage, phys. stat. sol. c \textbf{2}, 763 (2005); D.D. Solnyshkov, I.A. Shelykh, M.M. Glazov, G. Malpuech, T. Amand, P. Renucci, X. Marie, A.V. Kavokin, Semiconductors {\bf 41}, 1080 (2007).
\bibitem{COM02a}M. Combescot and O. Betbeder-Matibet, Europhys. Lett. {\bf 58}, 87 (2002).
\bibitem{COM02b}M. Combescot and O. Betbeder-Matibet, Europhys. Lett. {\bf 59}, 579 (2002). 
\bibitem{COM04} M. Combescot and O. Betbeder-Matibet, Phys. Rev. Lett. {\bf 93}, 016403 (2004).
\bibitem{COM07} M. Combescot and O. Betbeder-Matibet and R. Combescot, Phys. Rev. B {\bf 75}, 174305 (2007).
\bibitem{COM08} M. Combescot, O. Betbeder-Matibet, and F. Dubin, Phys. Rep. {\bf 463}, 215 (2008).
\bibitem{COM07b}M. Combescot, M. A. Dupertuis and O. Betbeder-Matibet, Europhys. Lett. {\bf 79}, 17001 (2007).
\bibitem{SAB01} M. Saba, C. Ciuti, J. Bloch, V. Thierry-Mieg, R. Andre, L. S. Dang, S. Kundermann, A. Mura, G. Bongiovanni, J. L. Staehli, and B. Deveaud,
Nature {\bf 414}, 731 (2001).
\bibitem{CIU01} C. Ciuti, P. Schwendimann, and A. Quattropani, Phys. Rev. B {\bf 63}, 041303(R) (2001).
\bibitem{CIU03} C. Ciuti, P. Schwendimann, and A. Quattropani, Semicond. Sci. Technol. {\bf 18}, S279 (2003).
\bibitem{ROC00} G. Rochat, C. Ciuti, V. Savona, C. Piermarocchi, A. Quattropani, and P. Schwendimann, Phys. Rev. B {\bf 61}, 13856 (2000).
\bibitem{CIU00} C. Ciuti, P. Schwendimann, B. Deveaud and A. Quattropani, Phys. Rev. B {\bf 62}, R4825 (2000).
\bibitem{ll3_eng} L.D. Landau and E.M. Lifshitz, \emph{Quantum Mechanics: Non-Relativistic Theory} vol. 3 (Butterworth-Heinemann, Oxford 1977).
\bibitem{averbuch86} P. G. Averbuch, J. Phys. A. {\bf 19}, 2325 (1986).
\bibitem{verhaar} B.J. Verhaar, J. P. H. W. van der Eijnde, M. A. J. Voermans, and M. M. J Schaffrath, J. Phys A. {\bf 17}, 595 (1984).
\bibitem{shumway06} J. Shumway, Physica E {\bf 32}, 273 (2006).
\bibitem{IVA04} A. L. Ivanov, P. Borri, W. Langbein, and U. Woggon, Phys. Rev. B {\bf 69}, 075312 (2004).
\end{thebibliography}
\end{document}